\documentclass[twoside]{article}
\usepackage[accepted]{aistats2024}
\usepackage[round]{natbib}

\usepackage[autonum,blackhypersetup]{shortex}

\begin{document}

\twocolumn[

\aistatstitle{Mixed variational flows for discrete variables}

\aistatsauthor{ Gian Carlo Diluvi \And Benjamin Bloem-Reddy \And  Trevor Campbell }

\aistatsaddress{Department of Statistics\\University of British Columbia  
\And  Department of Statistics\\University of British Columbia\\
\texttt{\{gian.diluvi, benbr, trevor\}@stat.ubc.ca}  
\And  Department of Statistics\\University of British Columbia} 
]

\begin{abstract}

Variational flows allow practitioners to learn 
complex continuous distributions,
but approximating \emph{discrete} distributions remains a challenge.
Current methodologies typically embed the discrete target in
a continuous space---usually via continuous relaxation
or dequantization---and then apply a continuous flow.
These approaches involve a surrogate target that
may not capture the original discrete target,
might have biased or unstable gradients, 
and can create a difficult optimization problem.
In this work,
we develop a variational flow family for discrete distributions
without any continuous embedding.
First, we develop a \emph{Measure-preserving And Discrete (MAD)} invertible
map that leaves the discrete target invariant, 
and then create a mixed variational flow (\emph{MAD Mix}) based on that map.
Our family provides access to i.i.d. sampling 
and density evaluation with virtually no tuning effort.
We also develop an extension to MAD Mix that handles joint
discrete and continuous models.
Our experiments suggest that MAD Mix produces
more reliable approximations than continuous-embedding flows
while requiring orders of magnitude less compute.

\end{abstract}

\section{INTRODUCTION}

The Bayesian statistical framework allows practitioners
to model complex relationships between variables of interest
and to incorporate expert knowledge as part of inference
in a principled way.
This has become crucial with the advent of heterogeneous data,
which is typically modeled using a mix of continuous and discrete latent variables.
One popular methodology for inference in Bayesian models is 
variational inference (VI) \citep{jordan1999vi,wainwright2008vi},
which involves finding a distribution
in a variational family of candidate distributions that minimizes a divergence to the posterior.
Distributions in the variational family usually enable both i.i.d.~draws 
and tractable density evaluation,
which allows practitioners to assess the quality of (and therefore optimize) 
the approximate distribution  by estimating, e.g., 
the evidence lower bound (ELBO) \citep{blei2017vi}.

State-of-the-art variational methods 
have been very successful in approximating \emph{continuous} distributions.
Of particular interest to this work are normalizing flows
\citep{tabak2013nfs,dinh2015nfs,rezende2015nfs,kobyzev2020nfs,papamakarios2021nfs},
which leverage repeated applications of 
flexible bijective transformations to construct highly expressive approximations.
Under mild conditions,
some normalizing flows are universal approximators of continuous distributions
when the number of repeated applications of the transformation
(i.e., the \emph{depth} of the flow) grows to infinity
\citep{huang2018universality,huang2020universality,kong2020universality,
zhang2020universality,lee2021universality}.
In recent work, \citet{xu2023mixflows} designed flow-based variational families
that have compute/accuracy trade-off theoretical guarantees,
and which circumvent the need to optimize the parameters of the flow.

In contrast with the continuous setting,
work on normalizing flows for approximating \emph{discrete} 
distributions has been more limited.
Bijections between discrete sets---i.e., permutations---can
only rearrange the probability masses among the discrete values
without changing their values \citep{papamakarios2021nfs}.
Recent work has addressed this issue by 
embedding the discrete target distribution in a continuous space in various ways,
and then approximating it with continuous flows.

One way to do this is to approximate the discrete
distribution of interest with a continuous relaxation
\citep{maddison2017concrete,jang2017concrete,
tran2019discflows,hoogeboom2019intflows}.
In this case, one constructs a surrogate continuous distribution
in the simplex parametrized by a temperature---at 
zero temperature, the approximation becomes the original discrete target.
But relaxing a discrete distribution to a continuous one introduces a trade-off between
the fidelity of the approximation and the difficulty of learning the flow:
a low temperature distribution will be very ``peaky''
near the vertices of the simplex, causing gradient instability.
Another continuous-embedding strategy is dequantization
\citep{uria2013dequantization,theis2016dequantization,
hoogeboom2020dequantization,nielsen2020survae,zhang2021ivpf,chen2022tessellation},
whereby one adds continuous noise to the atoms of the discrete distribution.
However, dequantization-based methodologies are incompatible with categorical data
since switching the labels results in a different dequantization.
This was partially addressed by \citet{hoogeboom2021argmaxflows},
who replaced the rounding operation used to 
quantize the continuous surrogate with an argmax.
However, Argmax flows still require careful---and expensive---tuning of the 
parameters of the flow.
Previous work has also considered transformations that 
update discrete states by thresholding a continuous neural network 
\citep{hoogeboom2019intflows,berg2021idf++,tomczak2021idf},
but this can bias the gradient estimates of the continuous flow.
Yet another option is to encode a discrete distribution
into a continuous distribution and optimize the encoder
\citep{ziegler2019latentflows,lippe2021catflows}.
But approximating a surrogate target density that
takes values in an inherently different space
introduces error even if the optimal approximation is eventually found.

In this work,
we develop a flow-based variational family
to approximate discrete distributions
without embedding them into a continuous space.
Our family is based on a new \emph{Measure-preserving And Discrete (MAD)} map
that leaves the discrete target invariant.
The key idea behind MAD is to augment the discrete target with uniform variables,
which we use to update each discrete variable via 
an inverse-CDF-like deterministic move.
We then use the MAD map as a building block in a mixed variational flow
(MixFlow) \citep{xu2023mixflows}, which averages over repeated applications of MAD.
We call the resulting variational family \emph{MAD Mix}.
Unlike the MixFlow instantiation in \citet{xu2023mixflows}, 
which assumes that the target distribution is continuous,
our family is specifically designed to learn 
discrete distributions---both ordinal and categorical.
We also show how to combine MAD with the discretized Hamiltonian
dynamics from \citet{xu2023mixflows} to approximate joint
continuous and discrete targets, e.g., mixture models.
Through multiple experiments,
we compare MAD Mix with several continuous-embedding normalizing flows,
with mean-field VI \citep{wainwright2008vi}, and with Gibbs sampling.
Our results show comparable sampling quality to Gibbs sampling,
but with the ability to evaluate the density of the approximation---and
therefore to assess its quality via the ELBO---as well as 
better training efficiency, stability, and 
approximation quality than dequantization, Argmax flows,
and Concrete normalizing flows.
\section{BACKGROUND}

Consider a target distribution $\pi$ on a set $\mcX$.
We assume that $x\in\mcX$ can contain both
discrete- and real-valued elements,
and that $\pi$ admits a density with respect to 
a product of Lebesgue and counting measures on 
their respective real-valued and discrete components.
We will use the same symbol to denote both a distribution and its density.
In the setting of Bayesian inference,
$\pi$ is a posterior distribution whose density we can only evaluate 
up to a normalizing constant,
$\pi(x)=p(x)/Z$, where $p$ is known but $Z=\int p$ is not.

In its most common form,
variational inference (VI) refers to approximating $\pi$
with an element $q^\star$ of a family of 
parametric variational approximations 
$\mcQ=\{q_\lambda\given\lambda\in\Lambda\}$.
Usually, $q^\star=q_{\lambda^\star}$ is chosen to minimize the 
Kullback--Leibler (KL) divergence \citep{kl}
between elements in $\mcQ$ and $\pi$:
\[\label{eq:vi}
    \lambda^\star
    &=\argmin_{\lambda\in\Lambda}\kl{q_\lambda}{\pi}\\
    &=\argmin_{\lambda\in\Lambda}
    \int_{\mcX}q_\lambda(x)\log\frac{q_\lambda(x)}{p(x)}\,\dee x.
\]
The normalizing constant $Z$ can be factored out of the KL divergence,
which results in the optimization problem in \cref{eq:vi} \citep{blei2017vi}.
Typically, $\mcQ$ is designed to allow both density evaluation 
and i.i.d. sampling from its elements.
This enables the use of 
stochastic gradient optimization algorithms to find 
a stationary point of \cref{eq:vi}
\citep[e.g.,][]{ranganath2014bbvi,kucukelbir2017advi}.

Normalizing flows
\citep{tabak2013nfs,rezende2015nfs,dinh2015nfs,
kobyzev2020nfs,papamakarios2021nfs}
are a common approach to design such a $\mcQ$.
Normalizing flows build an approximation $q_{\lambda}$
by applying a differentiable bijective map $T_\lambda$ 
that also has a differentiable inverse---i.e., a \emph{diffeomorphism}---to
a reference distribution $q_0$ on $\mcX$: $q_{\lambda}=T_\lambda q_0$.
Here, $T_\lambda q_0$ is the pushforward of $q_0$ under $T_\lambda$. 
In practice, $T_\lambda$ is built by composing multiple ``simple'' maps
$T_\lambda=T_{N,\lambda_N}\circ\cdots\circ T_{1,\lambda_1}$,
each with its own parameters $\lambda_n$.
If each $T_{n,\lambda_n}$ is invertible and differentiable
then so is the resulting $T_\lambda$.
Recent work has focused on designing maps $T_{n,\lambda_n}$ 
that satisfy two desiderata:
the simple maps are easy to evaluate, invert, and differentiate,
and the resulting family is highly expressive
(e.g., it is a universal approximator as in \citet{huang2018universality,
huang2020universality,kong2020universality,
zhang2020universality,lee2021universality}).
If $x$ is real-valued, the density of $q_{\lambda}$
can be written down using the determinant Jacobian of $T_\lambda$,
$J_\lambda(x)
:=\left|\grad_x T_\lambda(x)\right|$. 
Specifically, using the change-of-variables formula,
\[ \label{eq:nfs_density}
    q_{\lambda}(x)
    =\frac{q_0(T_\lambda^{-1}(x))}{J_\lambda(T_\lambda^{-1}(x))},\qquad
    x\in\mcX.
\]
If $x$ is discrete,
the change-of-variables formula does not contain a determinant Jacobian term
and so \cref{eq:nfs_density} becomes
$q_{\lambda}(x)=q_0(T_\lambda^{-1}(x))$.
However, bijections on discrete spaces are just permutations,
and so it is impossible to build expressive approximations in this setting.

Beyond issues with discrete variables, 
normalizing flows require practitioners to optimize the flow parameters 
$\lambda$ as in \cref{eq:vi}, 
which can introduce additional optimization-related problems.
This is because the optimization in \cref{eq:vi}
is the only means for the variational approximation $q_\lambda$
to adapt to the target $\pi$.
Recent work addresses this issue by constructing flows with diffeomorphisms 
that are also \emph{ergodic and measure-preserving} for $\pi$ \citep{xu2023mixflows}.
A map $T_\lambda$ is ergodic for $\pi$ if, when applied repeatedly,
it does not get ``stuck'' in any non-trivial regions of $\mcX$, i.e.,
if $T_\lambda(A)=A$ implies $\pi(A)$ is either 0 or 1
for any measurable $A\subseteq\mcX$.
$T_\lambda$ is measure-preserving for $\pi$
if it leaves the distribution of samples from $\pi$ invariant,
that is, if $X\sim\pi$ implies $T_\lambda(X)\sim\pi$.
A measure-preserving and ergodic map $T_\lambda$
has the property that averaging repeated applications of $T_\lambda$
will approximate expectations under $\pi$
\citep[][Corollary~11.2]{birkhoff1931ErgodicTheorem,eisner2015ErgodicTheorem}:
\bthm[\citet{birkhoff1931ErgodicTheorem}]\label{thm:ergodic}
If $T:\mcX\to\mcX$ is measure-preserving and ergodic for $\pi$, 
then for any $f\in L^1(\pi)$
\[
    \lim_{N\to\infty}\frac{1}{N}\sum_{n=0}^{N-1} f(T^n x)
    =\int_{\mcX}f\,\dee\pi,\qquad
    \pi\text{-a.e. }x\in\mcX.
\]
\ethm
Leveraging this result,
a mixed variational flow \citep{xu2023mixflows}, or \emph{MixFlow},
is a mixture of repeated applications of a
measure-preserving and ergodic $T_\lambda$.
Regardless of whether $x$ is real- or discrete-valued,
MixFlows have MCMC-like convergence guarantees
in that they converge to the target in total variation 
for any value of $\lambda$ \citep[][Theorems~4.1--4.2]{xu2023mixflows}:
\[\label{eq:mixflows}
    &q_{N,\lambda}:=\frac{1}{N}\sum_{n=0}^{N-1}T_\lambda^n q_0,\\
    &\lim_{N\to\infty}\tvd{q_{N,\lambda}}{\pi}=0,\quad
    \forall\lambda\in\Lambda.
\]
The density $q_{N,\lambda}$ for real-valued $x\in\mcX$
can be evaluated by backpropagating the flow:
\[\label{eq:mixflows_density}
    q_{N,\lambda}(x)
    =\frac{1}{N}\sum_{n=0}^{N-1}
    \frac{q_0(T_\lambda^{-n}(x))}
    {\prod_{j=1}^n J_\lambda(T_\lambda^{-j}(x))},
\]
with $J_\lambda(x)=\left|\grad_x T_\lambda(x)\right|$.
Furthermore, i.i.d. samples can be generated by repeatedly pushing samples 
from $q_0$ through $T_\lambda$:
\[
    n\sim\distUnif\{0,\dots,N-1\},\quad
    X_0\sim q_0,\quad
    T_\lambda^n(X_0)\sim q_{N,\lambda}.
\]
Although the theoretical guarantees for MixFlows hold for continuous and discrete $x$,
the instantiation provided by \citet{xu2023mixflows} only
applies to real-valued variables since it is based on Hamiltonian dynamics.
Furthermore, as with normalizing flows,
the density formula in \cref{eq:mixflows_density} is only valid for continuous $x$.
In the next section, we develop a measure-preserving map for discrete distributions
and then show how to construct a MixFlow based on it,
including extending the density formula from
\cref{eq:mixflows_density} to that setting.
\section{MEASURE-PRESERVING AND DISCRETE
MIXFLOWS (MAD MIX)} \label{sec:main_idea}

In this section,
we develop a novel measure-preserving bijection for discrete variables
that does not embed the underlying distribution in a continuous space
and that can be used in flow-based VI methodologies.
We call this map MAD
since it is Measure-preserving And Discrete.
The key idea behind MAD is to augment the target density with
a set of auxiliary uniform variables,
which we then use to update the discrete components.
\cref{fig:mad_viz} shows a single pass of the MAD map and
\cref{alg:mad} contains pseudocode to evaluate it.
We then construct a MixFlow based on the MAD map (\emph{MAD Mix})
and discuss how to extend MAD Mix to work with
joint discrete and continuous variables
by combining it with the prior work on MixFlows in \citet{xu2023mixflows}.

\subsection{Measure-preserving and discrete map}
To build intuition, 
we first consider a simplified example where 
we approximate a univariate discrete target $\pi$.
Without loss of generality, we assume $\mcX=\nats$.
In \cref{subsec:mad_extension},
we discuss how to use this simplification as a stepping stone
for the case where $\pi$ is multivariate.
We start by considering an augmented target density 
that contains a uniform variable $u\in[0,1]$:
\[
  \tdpi(x,u)=\pi(x)\ind_{[0,1]}(u).
\]
We use $u$ to sequentially update the value of $x$
with an inverse-CDF deterministic update.
Specifically, recall that if $X$ is a random variable with cdf $F$
and quantile function $Q(u)=\inf\{x\given F(x)\geq u\}$,
$u\in[0,1]$,
then the random variable $Q(U)$,
with $U\sim\distUnif[0,1]$, has cdf $F$
\citep[][Theorem~2.1]{devroye1986sampling}.
We leverage this fact and construct our map $T_{\mathrm{MAD}}$ 
by composing three steps.
First we will transform $u$ into a variable $\rho\in(0,1)$
that is in a direct inverse-CDF relationship with $x$.
Then we will update $\rho$ through an ergodic operator,
which in turn will induce a deterministic inverse-CDF update of $x$.
Finally, we transform $\rho$ back into $u$.
We detail each step below.

\begin{figure*}[t]
  \centering
  \includegraphics[width=\textwidth]{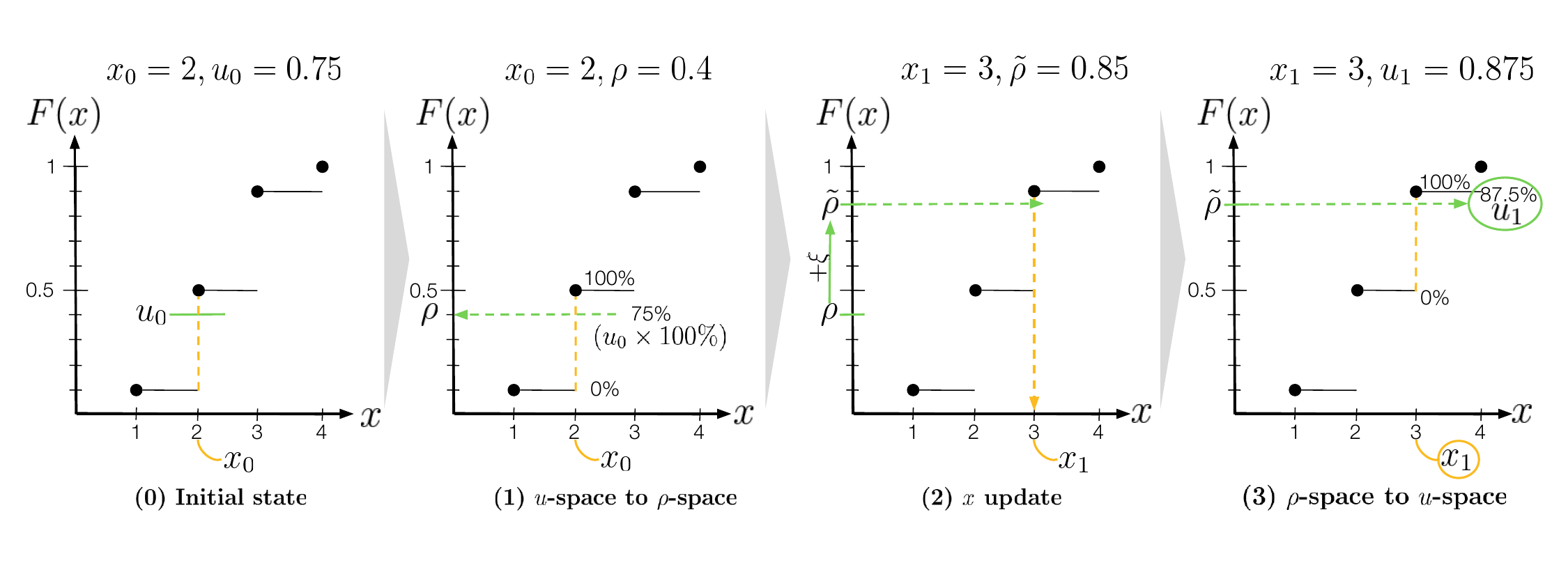}
  \caption{One application of the MAD map to the initial values
  $(x_0,u_0)=(2,0.75)$ with target probabilities 
  $\pi=\distCat(0.1,0.4,0.4,0.1)$.
  Each plot contains the CDF of $\pi$.
  In the first plot,
  $u_0$ represents the proportion of mass between
  $x_0=2$ and $1$.
  In the second plot, 
  $u_0$ is transformed into $\rho$---which
  indicates where in the CDF of $\pi$ the initial value $x_0$ lies at.
  Then, in the third plot, $\rho$ is shifted vertically by $\xi=0.45$
  to $\tdrho$, which produces a new value $x_1$ via the inverse-CDF trick.
  Finally, $\tdrho$ gets transformed into $u_1$,
  which represents the proportion of mass between $x_1=3$ and $2$.}
  \label{fig:mad_viz}
\end{figure*}

\paragraph{(1) Mapping uniform to $\rho$-space}
Let $F$ denote the CDF of $\pi$ and define $F(0)=0$.
Then we transform the uniform variable $u$ into $\rho$:
\[\label{eq:u_shift}
  \rho = F(x-1)+u\pi(x).
\]
While $u$ and $x$ are independent a priori,
this update introduces a dependence relationship between them by allowing
the uniform variable to switch between two interpretations:
$\rho\in[0,1]$ is the usual uniform $[0,1]$ variable in the inverse-CDF
method for drawing $x$,
while $u$ can be thought of as a proportion of the mass at $x$
and is independent of the value of $x$.
The first two panels of \cref{fig:mad_viz} show 
an example where $x=2$, $\pi(2)=0.4$, and $u=0.75$.
Here, $u$ indicates that $\rho$ lies 75\% of the way between $x=1$ and $x=2$,
and hence $\rho = F(1) + 0.75 \pi(2) = 0.1 + 0.75 \times 0.4 = 0.4$.
The Jacobian of this transformation w.r.t. $u$ is $\pi(x)$.

\paragraph{(2) State update}
We now do a shift in $\rho$-space:
\[
  \tdrho=\rho+\xi\mod1,
\]
where $\xi\in\reals$. 
If $\xi$ is irrational then this is 
an ergodic transformation for the uniform distribution;
in our experiments we used $\xi=\pi/16$.
The Jacobian of this transformation w.r.t. $\rho$ is 1 
(see \cref{lem:lcg_der}).
Since $\rho$ is marginally uniform $[0,1]$, 
the shift by $\xi$ preserves its distribution.
But note that $\rho$ and $x$ are jointly in an 
inverse-CDF relationship; so in order to preserve the 
joint uniform distribution on $\rho$ and inverse-CDF value $x$, 
we also need to update $x$ using the inverse-CDF method
described before:
\[\label{eq:quantile}
  x'=Q(\tdrho),\quad
  Q(p):=\min\{l\in\nats\given F(l)>p\},\,p\in[0,1].
\]
This transformation leaves the joint $(\rho, x)$ distribution invariant.
A similar idea is used by \citet{xu2023mixflows,murray2012detmcmc,neal2012permcmc}.

\paragraph{(3) Mapping back from $\rho$-space to $u$-space}
We finally map $\tdrho$ back to $u$-space:
\[\label{eq:u_correction}
  u'=\frac{\tdrho-F(x'-1)}{\pi(x')}.
\]
The Jacobian of this transformation w.r.t. $\tdrho$ is $1/\pi(x')$.
Because $\tdrho$ is marginally uniform $[0,1]$ 
and $x'$ is in an inverse-CDF relationship with it after steps (1) and (2),
this final step ensures that $u'$, $x'$ are independent 
and drawn from the augmented target $\tdpi(x,u)$. 
Therefore, steps (1)--(3) together leave $\tdpi(x,u)$ invariant.

\subsection{Multivariate MAD map} \label{subsec:mad_extension}
We now extend $T_{\mathrm{MAD}}$ to the case where $\pi$ 
has $M$ discrete variables.
Again without loss of generality, we encode these as $\mcX=\nats^M$.
Our extension of $T_{\mathrm{MAD}}$ to this setting
is inspired by the deterministic MCMC samplers in
\citet{neal2012permcmc,murray2012detmcmc} 
and the Gibbs samplers in 
\citet{geman1984gibbs,gelfand1990gibbs,neklyudov2021detgibbs}.
Specifically, $T_{\mathrm{MAD}}$ will mimic a single pass of a Gibbs sampler
targeting $\pi$, i.e.,
each iteration involves generating a sample from the 
full conditional distributions of $\pi$:
$\pi_m(x_m\given x_1,\dots,x_{m-1},x_{m+1},\dots,x_M)$
for $m=1,\dots,M$.
We achieve this by sequentially applying the univariate MAD map
to each full conditional.
For this purpose,
we introduce $M$ auxiliary uniform variables $u\in[0,1]^M$
to drive the updates of $x$ via deterministic inverse-CDF transforms
and consider an augmented target density:
\[
  \tdpi(x,u)=\pi(x)\ind_{[0,1]^M}(u).
\]
Given values of $(x,u)$,
$T_{\mathrm{MAD}}$ will sequentially update the $m$th entries
of $x$ and $u$, $(x_m,u_m)$, to $(x_m',u_m')$ given
the current values of $x$,
$(x_1',\dots,x_{m-1}',x_{m+1},\dots,x_M)$.
Specifically, we construct $T_{\mathrm{MAD}}=T_M\circ\cdots\circ T_1$
where each individual map $T_m$ is a pass of the MAD map
for univariate discrete distributions targeting
the augmented full conditional
$\tdpi_m(x_m,u_m)=\pi_m(x_m)\ind_{[0,1]}(u_m)$
(where we omit conditioning in the notation for brevity).
Note that $T_m$ only modifies $(x_m,u_m)$ for each $m$,
mimicking the sequential sampling from the full conditionals in Gibbs sampling,
where one conditions on the latest available value of each variable.
\cref{alg:mad} shows a single pass of $T_{\mathrm{MAD}}$.

\begin{algorithm}[t!]
  \caption{MAD map}
  \label{alg:mad}
  {
  \begin{algorithmic}[1]
  \Procedure{$T_{\mathrm{MAD}}$}{$x,u;\pi,\xi$}
  \State $J\gets1$
  \For{$m=1,\dots,M$}
  \State // \emph{Note: $\pi_m$, $F_m$, and $Q_m$ are for the}
  \State // \emph{augmented conditional distribution of $x_m$}
  \State // \emph{given the partially updated state}
  \State // \emph{$(x'_1, \dots, x'_{m-1}, x_{m+1}, \dots, x_M)$.}
  \State {\bf (1) Mapping uniform to $\rho$-space:}
  \State $\rho_m\gets F_m(x_m-1)+u_m\pi_m(x_m)$
  \State {\bf (2) State update:}
  \State $\tdrho_m\gets\rho+\xi\mod1$
  \State {\bf (3) Mapping back $\rho$-space to $u$-space:}
  \State $x_m'\gets Q_m(\tdrho_m)$
  \State $u_m'\gets\frac{\tdrho_m-F_m(x_m'-1)}{\pi_m(x_m')}$
  \State $J\gets J\cdot\pi_m(x_m)/\pi_m(x_m')$ 
  \EndFor
  \State\Return $x',u',J$
  \EndProcedure
  \end{algorithmic}
  }
  \end{algorithm}

\subsection{Theoretical properties of the MAD map} \label{subsec:theory}

Now we show that our construction of $T_{\mathrm{MAD}}$ 
has a tractable inverse and that it leaves the augmented target $\tdpi$ invariant.
We also show how to compute the density of pushforwards through $T_{\mathrm{MAD}}$.

\paragraph{$T_{\mathrm{MAD}}$ is invertible}
Each map $T_m$ is invertible:
computing $T_m^{-1}$ is equivalent to 
evaluating $T_m$ with the inverse shift $-\xi$.
Hence evaluating $T_{\mathrm{MAD}}^{-1}$ amounts to 
propagating each flow $T_m$ \emph{forward} with a negative shift
and in reverse order (i.e., starting from $T_M$ since
$T_{\mathrm{MAD}}^{-1}=T_1^{-1}\circ\dots\circ T_M^{-1}$).

\paragraph{Density of pushforward under $T_{\mathrm{MAD}}$}
Care is needed since the standard change-of-variables formula 
only applies when all the variables are either real or discrete.
In our setting, $x$ is discrete and $u$ is real,
so we develop a change of variables analogue for this setting in \cref{prop:cov}.
Denote by $T_{\mathrm{d}}(x,u)$ and $T_{\mathrm{c}}(x,u)$
the discrete and continuous components of $T_{\mathrm{MAD}}$, respectively:
$T_{\mathrm{MAD}}(x,u)=(x',u'):=(T_{\mathrm{d}}(x,u),T_{\mathrm{c}}(x,u))$.
By \cref{prop:cov}, for $(X, U)\distas g$ from some base distribution $g$
and $(X',U')=T_{\mathrm{MAD}}(X,U)$, the density of $(X',U')$ is
\[\label{eq:cov_T}
  &\frac{g(T_{\mathrm{MAD}}^{-1}(x',u'))}
  {J_{\mathrm{c}}(T_\mathrm{MAD}^{-1}(x',u'))},\\
  &J_{\mathrm{c}}(x,u)
  =\left|\grad_u T_{\mathrm{c}}(x,u)\right|
  =\prod_{m=1}^M \frac{\pi_m(x_m)}{\pi_m(x_m')},
\]
where $J_{\mathrm{c}}(x,u)$
is the product of the Jacobians from Steps (1)--(3)
since $T_m$ only affects $(x_m,u_m)$.

\paragraph{$T_{\mathrm{MAD}}$ is measure-preserving for $\tdpi$}
We now show in three steps that $T_{\mathrm{MAD}}=T_M\circ\dots\circ T_1$ 
is measure-preserving for $\tdpi$.
First we show in \cref{prop:fc_mp} that each $T_m$ is
measure-preserving for the corresponding full conditional $\tdpi_m$.
Then we show in \cref{prop:slice_mp} that a map that
only modifies some values of its input \emph{and} 
is measure-preserving for the full conditional of those values 
is also measure-preserving for the joint distribution.
The result then follows from the fact that a composition of
measure-preserving maps for $\tdpi$ is also measure-preserving for $\tdpi$.

\bprop\label{prop:fc_mp}
  $T_m \tdpi_m=\tdpi_m$.
\eprop

\bprop\label{prop:slice_mp}
  Let $\pi(\dee x_1,\dee x_2)$ be a measure defined on
  $\mcX_1\times\mcX_2\subseteq\reals^{d_1}\times\reals^{d_2}$.
  with disintegration
  $\pi(\dee x_1,\dee x_2)=\pi_1(\dee x_1)\pi_{2|1}(\dee x_2,x_1)$.
  Let $T_{x_1}(x_2)$ be a $\pi_{2|1}$-measure-preserving transformation.
  Then $T(x_1,x_2):=(x_1,T_{x_1}(x_2))$ is $\pi$-measure-preserving.
\eprop

The proof of \cref{prop:fc_mp} is based on 
the argument in \citet{murray2012detmcmc} and amounts to using 
the change-of-variables formula in \cref{prop:cov}
to directly compute the density of the pushforward,
which coincides with $\tdpi_m$.
\cref{prop:slice_mp} is proved by disintegrating the base measure
into its full conditionals, applying the measure-preserving property,
and then integrating back w.r.t. the joint measure.
Complete proofs are in \cref{sec:proofs}.

\subsection{Approximating discrete distributions with MAD Mix}
Now we show how to use $T_{\mathrm{MAD}}$
to approximate discrete distributions.
Let $q_0$ be a reference distribution on $\mcX\times[0,1]^M$.
We construct a MixFlow \citep{xu2023mixflows} $q_N$ based on the MAD map,
i.e., a mixture of repeated applications of $T_{\mathrm{MAD}}$.
We can express the density of the approximation applying the
formula for the density under a single pass of $T_{\mathrm{MAD}}$ in \cref{eq:cov_T}
to each element in the mixture:
\[\label{eq:qN_density}
  q_N(x,u)=
  \frac{1}{N}\sum_{n=0}^{N-1}
    \frac{q_0(T_{\mathrm{MAD}}^{-n}(x,u))}
    {\prod_{j=1}^n J_{\mathrm{c}}(T_{\mathrm{MAD}}^{-j}(x,u))}.
\]
The density \cref{eq:qN_density} can be computed efficiently
by caching the determinant Jacobians during backpropagation,
as shown in \cref{alg:madam} 
(which is an instantiation of \citet[][Algorithm~2]{xu2023mixflows}).
Sampling from $q_N$ can be done by first drawing an $n \distas \distUnif\{0,\dots,N-1\}$,
then drawing $(X,U)\sim q_0$, and finally pushing $(X,U)$ 
through $T_{\mathrm{MAD}}^n$.
Our variational family has no parameters other than $N$,
so that tracking the ELBO is done to assess the quality of the approximation
rather than to optimize any parameters---a costly operation needed 
in other variational methodologies.

\begin{algorithm}[t!]
  \caption{MAD Mix logdensity}
  \label{alg:madam}
  {
  \begin{algorithmic}[1]
  \Procedure{$\log q_N$}{$x,u;N,q_0,\pi,\xi$}
  \State $L\gets0$
  \State $w_0\gets\log q_0(x,u)$
  \For{$n=1,\dots,N-1$}
  \State $x,u,J\gets T_{\mathrm{MAD}}^{-1}(x,u;\pi,\xi)$ (see \cref{alg:mad})
  \State $L\gets L+\log J$
  \State $w_n\gets\log q_0(x,u)-L$
  \EndFor
  \State\Return $\mathtt{LogSumExp}(w_0,\dots,w_{N-1})-\log N$
  \EndProcedure
  \end{algorithmic}
  }
  \end{algorithm}

\subsection{Approximating joint discrete and continuous distributions 
with MAD Mix} \label{subsec:madmix_extension}
Many practical situations with discrete variables also contain continuous variables.
We show how to combine our map $T_{\mathrm{MAD}}$ with 
the instantiation for continuous variables based on 
uncorrected Hamiltonian dynamics in \citet{xu2023mixflows}.
Suppose we have a target density $\pi(x_{\mathrm{c}},x_{\mathrm{d}})$
on $\mcX_{\mathrm{c}}\times\mcX_{\mathrm{d}}$, 
where $\mcX_{\mathrm{c}}\subseteq\reals^{M_{\mathrm{c}}}$
is a space of continuous variables
and $\mcX_\mathrm{d}\subseteq\nats^{M_{\mathrm{d}}}$ is a space of discrete variables.
We consider the following augmented target density on 
$\mcX_{\mathrm{c}}\times\reals^{M_{\mathrm{c}}}\times[0,1]
\times\mcX_{\mathrm{d}}\times[0,1]^{M_{\mathrm{d}}}$:
\[
  \tdpi(x_{\mathrm{c}},m,u_{\mathrm{c}},x_{\mathrm{d}},u_{\mathrm{d}})
  =\pi(x_{\mathrm{c}},x_{\mathrm{d}})r(m)
  \ind_{[0,1]^{M_{\mathrm{d}}+1}}(u_{\mathrm{c}},u_{\mathrm{d}}).
\]
Above, we introduced $M_{d}+1$ uniform variables and
$M_{c}$ \emph{momentum} variables $m$ with density
$r(m)=\prod_i r_0(m_i)$, where $r_0$ is a base distribution
(we used Laplace in our experiments).
\citet{xu2023mixflows} describe a measure-preserving map $H$
that mimics uncorrected Hamiltonian dynamics.
We combine their map and ours into a mixed map $\hT$
that sequentially updates all variables:
$\hT=\hT_{\mathrm{MAD}}\circ\hH
:=(\mathrm{Id},T_{\mathrm{MAD}})\circ(H,\mathrm{Id})$,
where $\mathrm{Id}$ is an identity map of the appropriate dimension
and we use a hat to denote extension by the identity map.
That is, $\hT$ is defined in two steps via
\[
  (x_{\mathrm{c}},m,u_{\mathrm{c}},
  x_{\mathrm{d}},u_{\mathrm{d}})
  &\overset{\hH}{\longmapsto}
  (x_{\mathrm{c}}',m',u_{\mathrm{c}}',
  x_{\mathrm{d}},u_{\mathrm{d}})\\
  &\overset{\hT_{\mathrm{MAD}}}{\longmapsto}
  (x_{\mathrm{c}}',m',u_{\mathrm{c}}',
  x_{\mathrm{d}}',u_{\mathrm{d}}').
\]
Since the continuous and discrete maps are measure-preserving
for their respective full conditionals,
it follows from \cref{prop:slice_mp} that they are also measure-preserving for $\tdpi$.
Therefore, $\hT$ is also measure-preserving for $\tdpi$ since it
composes (the identity-extended) $H$ and $T_{\mathrm{MAD}}$.

Let $\tdq_N=N^{-1}\sum_{n=0}^{N-1}\hT^n q_0$
with a reference $q_0$ over the augmented space.
Then the density $\tdq_N({\bf x})$ with 
${\bf x}=(x_{\mathrm{c}},m,u_{\mathrm{c}},x_{\mathrm{d}},u_{\mathrm{d}})$
can be evaluated by inverting $\hT$ and multiplying the two Jacobians:
\[
  &\frac{1}{N}\sum_{n=0}^{N-1}
  \frac{q_0(\hT^{-1}({\bf x}))}
  {\prod_{j=1}^n \hJ_{\mathrm{c}}(\hT_{\mathrm{MAD}}^{-1}
  \circ\hT^{-j+1}({\bf x}))
  \hJ_{\mathrm{Ham}}(\hT^{-j}({\bf x}))},
\]
with $\hJ_{\mathrm{c}}({\bf x})$ the identity-extended version
of $J_{\mathrm{c}}(x)$ and $\hJ_{\mathrm{Ham}}$ the
extended Jacobian of $H$ \citep[see][]{xu2023mixflows}.

\subsection{Comparison with existing deterministic MCMC methods}

The MAD map is inspired by 
deterministic Markov chain Monte Carlo (MCMC) samplers
\citep{murray2012detmcmc,neal2012permcmc,chen2016herdedgibbs,
wolf2020herdedGibbs,neklyudov2020imcmc,neklyudov2021detgibbs,
versteeg2021nonnewtonianhmc,seljak2022detLangevin,neklyudov2022orbital},
which provide a first step in designing 
tractable ergodic and measure-preserving maps.
Most of these, however, are designed with sampling as the only goal
and thus do not have a tractable inverse map.
Our work is the first to keep track of the inverse map
and use the resulting deterministic MCMC operator
within a flow-based variational family.

Of most relevance to this work are the deterministic MCMC operators developed by
\citet{neal2012permcmc,murray2012detmcmc,neklyudov2021detgibbs},
and particularly the discrete map from \citet{murray2012detmcmc}.
However, instead of switching their uniform variables between two spaces 
(i.e., $u$-space and $\rho$-space in our case) to drive the discrete updates,
\citet{murray2012detmcmc} only work in $\rho$-space and impose an additional restriction
on the last uniform update to ensure measure invariance.
Another major difference with our work is that
they propose using the same update for continuous variables.
This requires access to the reverse MCMC operator,
which limits the applicability of their methodology.
Instead, we extend MixFlows to learn jointly discrete and continuous
target distributions with \cref{prop:fc_mp},
and then develop a MixFlow instantiation for this setting based on 
our new MAD map and the discretized uncorrected Hamiltonian
dynamics from \citet{xu2023mixflows}.

\section{EXPERIMENTS} \label{sec:experiments}

In this section,
we compare the performance of MAD Mix against 
Gibbs sampling \citep{geman1984gibbs,gelfand1990gibbs},
mean-field VI \citep{wainwright2008vi},
dequantization \citep{uria2013dequantization,theis2016dequantization},
Argmax flows \citep{hoogeboom2021argmaxflows},
and Concrete relaxations \citep{maddison2017concrete,jang2017concrete}.
For the three continuous-embedding flow-based methods,
we implemented a Real NVP normalizing flow \citep{dinh2017realnvp}.
The discrete components were either dequantized,
argmaxed, or approximated with Concrete relaxations.
We consider five discrete experiments and four 
joint discrete and continuous experiments
(all with real-world data).
For each experiment,
we performed a wide architecture search
with different settings for the normalizing flows.
See \cref{sec:implementation} for more details.
All experiments were conducted on a machine with 
an Apple M1 chip and 16GB of RAM.
Code to reproduce experiments is available at
\url{https://github.com/giankdiluvi/madmix}.

\begin{figure*}[t!]
    \centering
    \begin{subfigure}{0.37\linewidth}
        \centering
        \includegraphics[width=\linewidth]{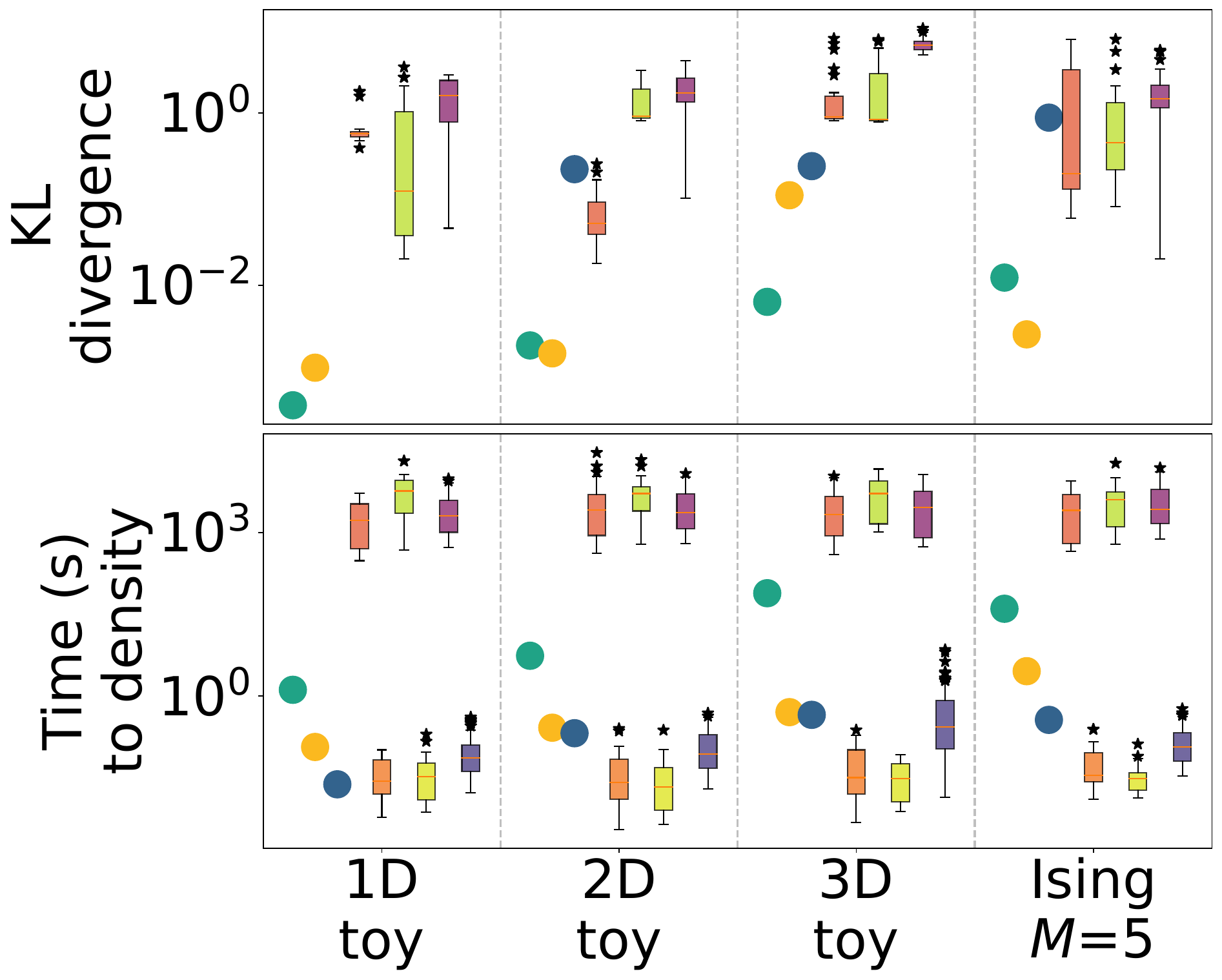}
        \caption{Normalized target}
        \label{fig:norm_summary}
    \end{subfigure}%
    \begin{subfigure}{0.61\linewidth}
        \centering
        \includegraphics[width=\linewidth]{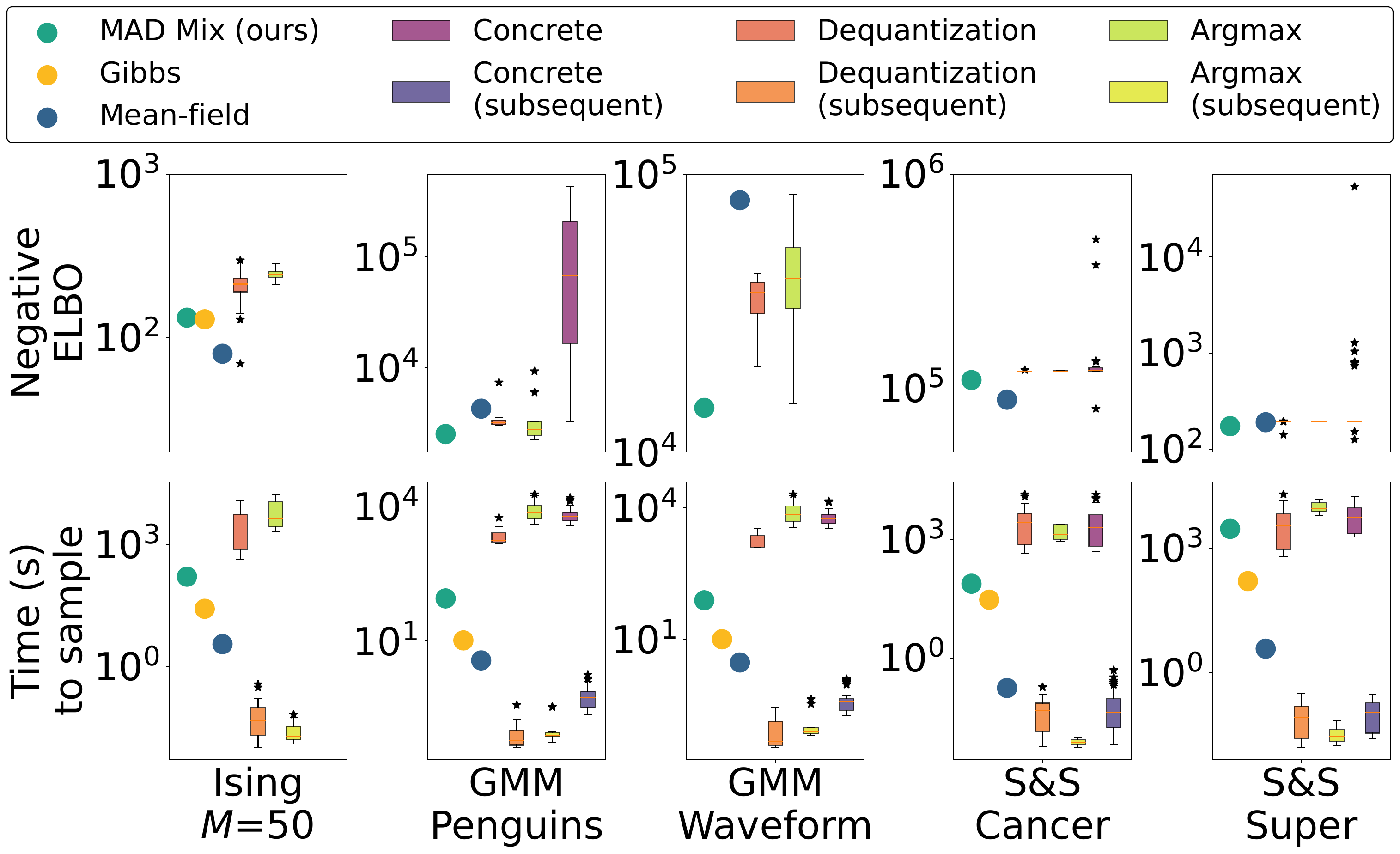}
        \caption{Unnormalized target}
        \label{fig:unnorm_summary}
    \end{subfigure}
    \caption{
    Summary of experiments.
    In \cref{fig:norm_summary}, the 
    normalizing constant $Z$ of the target density is known
    while in \cref{fig:unnorm_summary} $Z$ is not tractable.
    The boxplots for continuous-embedding flows represent the 
    search over different architecture settings.
    (Top row): KL divergence (\cref{fig:norm_summary})
    and negative ELBO (\cref{fig:unnorm_summary})
    from approximation to target distribution.
    Lower is better.
    (Bottom row): Compute time (seconds)
    to evaluate or estimate the density (\cref{fig:norm_summary})
    or to generate a sample (\cref{fig:unnorm_summary}).
    The second set of boxplots for
    continuous-embedding flows show the time
    to evaluate a subsequent density point after training.
    Missing values indicate either that 
    the algorithm cannot be used for that task or 
    that it was too computationally unstable to produce results,
    except for 1D mean-field VI
    which produces exact results ($\mathrm{KL}=0$).
    Colors are shared across figures and $x$-axes across columns.}
    \label{fig:summary}
\end{figure*}

\cref{fig:summary} presents a summary of our experiments.
MAD Mix obtains good approximations across the board at 
a fraction of the compute cost of dequantization,
Argmax flows, and Concrete-relaxed flows.
The time plot includes the time necessary to 
train all the flows in the architecture search.
Since training the flow is a computational bottleneck,
we also show a separate set of boxplots for the time required to
evaluate the density of the approximation after training.
This shows that,
although density evaluation is computationally cheap given a trained flow,
optimization can be costly and should be accounted for in the compute budget.
We also found that Concrete-relaxed flows were prone to numerical instability and
could not always be inverted for density evaluation in more complex examples,
as shown by their omission in \cref{fig:unnorm_summary}.
This behaviour has been documented in the past 
\citep[e.g.,][Sec.~3.7]{dinh2017realnvp}.
Gibbs sampling and mean-field VI require less compute time to 
sample or to evaluate a density than MAD Mix; see
the bottom plots of \cref{fig:norm_summary,fig:unnorm_summary}.
However, the former struggles to produce 
high-quality approximations in complex examples
and the latter does not provide access to the density of the approximation.
This is shown in \cref{fig:unnorm_summary},
where the ELBO could not be estimated for Gibbs sampling.

\subsection{Discrete toy examples} \label{subsec:discrete_toy}

First, we consider three discrete toy examples:
a 1D distribution on $\{1,\dots,10\}$,
a 2D distribution on $\{1,\dots,4\}\times\{1,\dots,5\}$,
and a 3D distribution on $\{1,\cdots,10\}^3$,
all generated randomly.
In all cases, MAD Mix produces a high-fidelity approximation
of the target distribution as seen in 
\cref{fig:norm_summary,fig:discrete_toy_examples}.
In contrast, continous-embedding flows consistently require more compute
and produce, on average, worse approximations.
Concrete-relaxed and Argmax flows produce good global approximations
but generally fail to recover the local shape.
This results in small but not negligible KL to the target,
as seen in \cref{fig:summary}.
Dequantization generally produces better approximations than Concrete,
but it failed to properly capture the shape of the simplest, 
1D toy example (see \cref{fig:onedim_prbs}).
This behaviour was consistent across all architecture configurations,
and a visual inspection of the loss traceplots suggests that
the optimizer converged in all cases.
Argmax flows show a similar behaviour in the 2D case (see \cref{fig:twodim_prbs1}).
Mean-field VI is computationally cheaper than MAD Mix
but in the multivariate examples it consistently produced worse approximations, i.e.,
with higher KL to the target.
Since the MAD map mimics a pass of a Gibbs sampler,
Gibbs sampling performs comparably to MAD Mix.
While Gibbs sampling is also cheaper due to not having to keep track of the 
auxiliary uniform variables,
it does not provide access to a density; the KL was estimated with an empirical PMF.
In more complex examples, this is not possible
and we cannot assess Gibbs' performance.

\subsection{Ising model}

Next, we consider an Ising model on $\mcX=\{-1,+1\}^M$ with log-PMF
$\log\pi(x)=\beta\sum_{m=1}^{M-1} x_m x_{m+1}-\log Z$,
where $\beta>0$ is the inverse temperature and
$M$ controls the dimension of the problem.
The normalizing constant $Z$ involves a sum over $2^M$ terms
(equal to the dimension of the latent space)
but the full conditionals $\pi_m$ can be calculated in closed form
for any $M$.
We considered a small $M$ and a large $M$ setting,
with dimensions $M=5$ and $M=50$
and inverse temperatures $\beta=1$ and $\beta=5$, respectively.
In the $M=5$ case,
the normalizing constant can be computed.
All methods except mean-field 
(due to the $M$ particles being treated as independent)
produce high-fidelity approximations; see \cref{fig:ising_prbs}.
However, only a few continuous-embedding flow architectures resulted in
small KL divergences, as seen in \cref{fig:norm_summary}.
The implementation in PyTorch \citep{paszke2019pytorch} of Concrete relaxations
requires access to all the (possibly unnormalized) probabilities.
Since allocating a vector of size $2^{50}$ is not possible,
we were unable to fit a Concrete-relaxed flow in the $M=50$ setting.
As seen in \cref{fig:unnorm_summary},
MAD Mix, dequantization, Argmax flows,
mean-field VI, and Gibbs sampling all perform comparably.
The ELBO estimates for Gibbs sampling, dequantization,
and Argmax flows were obtained via empirical
frequency PMF estimators
(the first since Gibbs does not provide access to densities,
the other two since one needs to approximate an $M$-dimensional integral
to estimate the PMF, which is not feasible for large $M$).
In contrast, both MAD Mix and mean-field VI provide access to exact PMFs.

\subsection{Gaussian mixture model}

To showcase a joint continuous and discrete example,
we considered a Gaussian mixture model (GMM)
with likelihood $\sum_{k=1}^K w_k \phi_{\mu_k,\Sigma_k}$,
where $w\in\Delta^{K-1}$ and $\phi_{\mu,\Sigma}$
is a Gaussian distribution with mean $\mu$ and covariance matrix $\Sigma$.
We fit this model to two datasets:
the Palmer penguins dataset\footnote{Available at 
\url{https://github.com/mcnakhaee/palmerpenguins}.}
and the waveform dataset.\footnote{Available at
\url{https://hastie.su.domains/ElemStatLearn/datasets/waveform.train};
see \citep[][p.~451]{hastie2009statlearning}.}
The former has 333 observations of four different
measurements of three species of penguins and
the latter contains 300 simulated observations
of 21 measurements of three classes.
Following \citet{hastie2009statlearning},
we use the first two principal components of the measurements as the observations.
In both examples, we perform inference
over the labels, the weights, and the measurement means and covariance matrices
of each species (penguins) and class (waveform),
for total latent space dimensions of 1044 for the penguins data
and 918 for the waveform data.
See \cref{sec:gmm_fc} for specific modeling details.
We fit the mixed discrete-continuous variant of MAD Mix
described in \cref{subsec:madmix_extension}.
\cref{fig:unnorm_summary} shows the results of both data sets.
Gibbs cannot produce an estimate of the ELBO and so we are
unable to assess its accuracy.
We also noticed that most of the architecture configurations
for the three continuous-embedding methods
resulted in gradient overflow.
Furthermore, the few Concrete flows in the waveform data set 
that were optimized produced covariance matrices 
whose diagonal was numerically zero.
We therefore could not evaluate the target density to then estimate the ELBO.
From \cref{fig:unnorm_summary},
MAD Mix produces approximations with a higher ELBO than
mean-field VI, dequantization, Argmax flows, and Concrete-relaxed flows.

\subsection{Spike-and-Slab model}

Finally, we considered a sparse Bayesian regression experiment
where we modeled $N$ observations $y_n\sim\distNorm(\beta^\top x_n, \sigma^2)$
and placed a spike-and-slab prior on the $P$ regression coefficients:
$\beta_p\sim (1-\gamma_p) \delta_0 + \gamma_p \distNorm(0,\nu^2)$
for all $p$.
We performed inference on the coefficients $\beta_p$, 
the binary variables $\gamma_p\in\{0,1\}$
that indicate whether $\beta_p=0$ or not,
and the variances $\sigma^2,\nu^2$.
See \cref{sec:sas_details} for more modeling details.
We fit this model to two datasets:
a pancreatic cancer dataset\footnote{Available 
at \url{https://hastie.su.domains/ElemStatLearn/datasets/prostate.data}.}
with $N=97$ and $P=8$
and a superconductivity dataset\footnote{Available 
at \url{https://archive.ics.uci.edu/dataset/464/superconductivty+data}.}
with $N=100$ (subsampled) and $P=81$.
The latent space dimensions are 27 for the cancer data set
and 246 for the superconductivity data set.
From \cref{fig:unnorm_summary},
all algorithms perform comparably.
As in the GMM experiment,
most of the optimization routines to train the
continuous-embedding flows resulted in gradient overflow.
This highlights the importance of doing an architecture search.
In contrast, MAD Mix and mean-field VI were only run once
and produced results comparable to the best continuous-embedding flows.
Gibbs sampling was also only run once, but as before we cannot compute the ELBO;
we expect it to behave similarly to MAD Mix.
Concrete-relaxed flows provide the best approximation in both data sets
as a result of a single neural network parameter configuration.
However, they also consistently produced very poor approximations in both cases.

\section{CONCLUSION} \label{sec:conclusion}

In this work we introduced \emph{MAD Mix}, a new variational family
to learn discrete distributions without embedding them in continuous spaces.
MAD Mix consists of a mixture of repeated applications of a novel
\emph{Measure-preserving And Discrete (MAD)} map 
that generalizes those used by deterministic MCMC samplers.
Our experiments show that MAD Mix produces high-fidelity approximations
at a fraction of the compute cost (including training)
than those obtained from continuous-embedding normalizing flows.
Since MAD Mix mimics sequentially sampling from the target's full conditionals,
the quality of the approximation in the discrete setting
will depend on the mixing properties of Gibbs sampling.
Future work can explore new measure-preserving and ergodic maps that
lead to flow-based families like MAD Mix but with better mixing.

\section*{Acknowledgments}

The authors gratefully acknowledge the support of the
National Sciences and Engineering Research Council of Canada (NSERC), 
specifically RGPIN2020-04995, RGPAS-2020-00095, DGECR2020-00343, and RGPIN-2019-03962,
as well as a UBC Four Year Doctoral Fellowship.
This research was supported in part through computational resources and services
provided by Advanced Research Computing at the University of British Columbia.

\bibliographystyle{abbrvnat}
\bibliography{ref}

\clearpage
\onecolumn
\appendix
\section{IMPLEMENTATION DETAILS} \label{sec:implementation}

\paragraph{MAD Mix}
There are two tunable parameters in MAD Mix:
the vertical shift $\xi$ in Step (2) and
the number $N$ of repeated applications of $T_{\mathrm{MAD}}$.
Following \citet{xu2023mixflows}, 
we fixed $\xi=\pi/16$ and found that it worked well in all our cases.
We chose $N$ to achieve a desirable tradeoff between
the accuracy of the approximation $q_N$  (measured by the ELBO) and
the time it took to evaluate the density of or generate samples from $q_N$.
We found that $N$ in the order of $10^2$ was sufficient for all of our experiments,
and specifically we set $N=500$ for the univariate and bivariate toy examples,
$N=100$ for the 3D toy example,
$N=1000$ for the low-dimensional Ising example
and $N=500$ for the high-dimensional one,
$N=100$ for both Gaussian mixture models,
and $N=500$ for both Spike-and-Slab examples.
$N$ can be thought of as the number of Gibbs sampling steps
being averaged over (since the MAD map mimics a deterministic Gibbs sampler).

\paragraph{Normalizing flow architecture}
We conducted a search over 144 different setting configurations
for Concrete-relaxed normalizing flows
and 36 different settings for Argmax flows and dequantization.
Our motivation was to reflect the effort of tuning 
continuous-embedding flows in practice
by searching over reasonable configuration attempts.

For our base architecture,
we considered a Real NVP normalizing flow \citep{dinh2017realnvp},
which has been shown to provide highly expressive approximations 
to continuous distributions.
Each pass of a Real NVP flow transforms $x\mapsto x'$
and is constructed by scaling and translating only some of the inputs:
\[ \label{eq:realnvp_update}
    x=[x_a,x_b],\quad
    x_a'=\exp(s(x_b;\psi_s))\odot x_a+t(x_b;\psi_t),\quad
    x':=[x_a',x_b],
\]
where $s$ and $t$ are neural networks
that depend on $x_b$ and parameters $\psi=(\psi_s,\psi_t)$
and $\odot$ indicates element-wise multiplication.
We defined $s$ and $t$ by composing
three applications of a single-layer linear feed forward neural network
with leaky ReLU activation functions in between.
For the scale transformation $s$,
we added a hyperbolic tangent layer too:
\[
    t(x_b;\psi_t)
    &=(\mathrm{Linear}\circ
    \mathrm{ReLU}\circ
    \mathrm{Linear}\circ
    \mathrm{ReLU}\circ
    \mathrm{Linear})(x_b;\psi_t),\\
    s(x_b;\psi_s)
    &=(\tanh\circ
    \mathrm{Linear}\circ
    \mathrm{ReLU}\circ
    \mathrm{Linear}\circ
    \mathrm{ReLU}\circ
    \mathrm{Linear})(x_b;\psi_s).
\]
The initial and final widths of $s$ and $t$
(i.e., the first and last Linear layers in $t$
and the first Linear and the hyperbolic tangent layers in $s$)
were chosen to match the dimension of $x$.
For example, in the 1D toy discrete example, 
the dimension of the Concrete-relaxed $x$ is 10 
due to one-hot encoding because 
the underlying discrete random variable takes values in $\{1,\dots,10\}$.
For the intermediate layers, we considered four different widths:
32, 64, 128, and 256.

These steps define a single pass of the Real NVP map,
but in practice it is necessary to do multiple passes
(the number of which is the \emph{depth} of the flow).
We considered three different depths: 10, 50, and 100.
To increase the expressiveness of the flow,
we alternated between updating the first half and the last half
of the inputs in each pass (as recommended by \citet{dinh2017realnvp}).
That is, on odd passes $x_a$ in \cref{eq:realnvp_update}
would correspond to the first half of $x$ and 
on even passes to the last half of $x$.

Additionally,
for Concrete-relaxed flows we considered
multiple possible values of the temperature parameter $\lambda$
that controls the fidelity of the relaxation:
when $\lambda\to0$, the relaxed distribution converges to the 
original discrete distribution,
but this in turn causes the relaxation to be ``peaky'' near the vertices
of the simplex and therefore gradients are numerically unstable.
Based on the discussion in \citet[][Appendix~C.4]{maddison2017concrete},
we considered four different temperatures: 0.1, 0.5, 1, and 5.
This range covers different approximation quality/computational stability
tradeoff regimes
and covers the optimal values found by \citet{maddison2017concrete}
in their experiments.
For the multivariate distributions,
we instead relaxed a ``flattened'' target distribution with dimension equal
to the product of the dimensions of each variable.
E.g., we mapped the bivariate toy example 
with values in $\{1,\dots,5\}\times\{1,\dots,4\}$ 
to a univariate distribution with values in $\{1,\dots,20\}$.

We implemented the Real NVP normalizing flow
in PyTorch \citep{paszke2019pytorch}.
In practice, we used a log-transformed Concrete approximation
as suggested in \citet[][Appendix~C.3]{maddison2017concrete},
which we found to be more numerically stable.
We learned the parameters of the flow by running Adam \citep{kingma2015adam}
for 10,000 iterations.
We considered three different learning rates: 
$10^{-5}$, $10^{-3}$, and $10^{-1}$.

We optimized all flows in parallel
using Sockeye \citep{sockeye}, 
a high-performance computing platform at our institution.

\section{PMF OF APPROXIMATIONS OF TOY EXAMPLES} \label{sec:discrete_toy_fig}

\begin{figure}[!ht]
    \centering
    \begin{subfigure}{0.45\linewidth}
        \centering
        \includegraphics[width=0.9\linewidth]{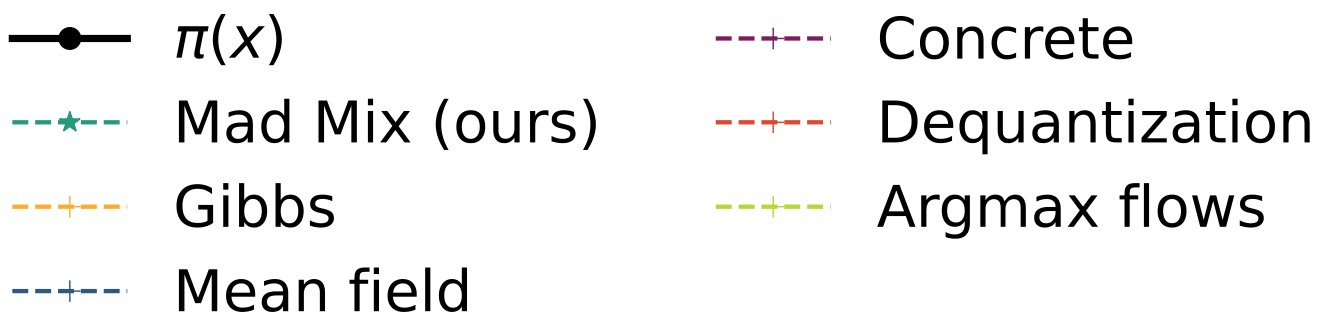}
        \vspace*{1.4cm}
        \caption*{}
        \label{fig:prbs_legend}
    \end{subfigure}%
    \begin{subfigure}{0.45\linewidth}
        \centering
        \includegraphics[width=\linewidth]{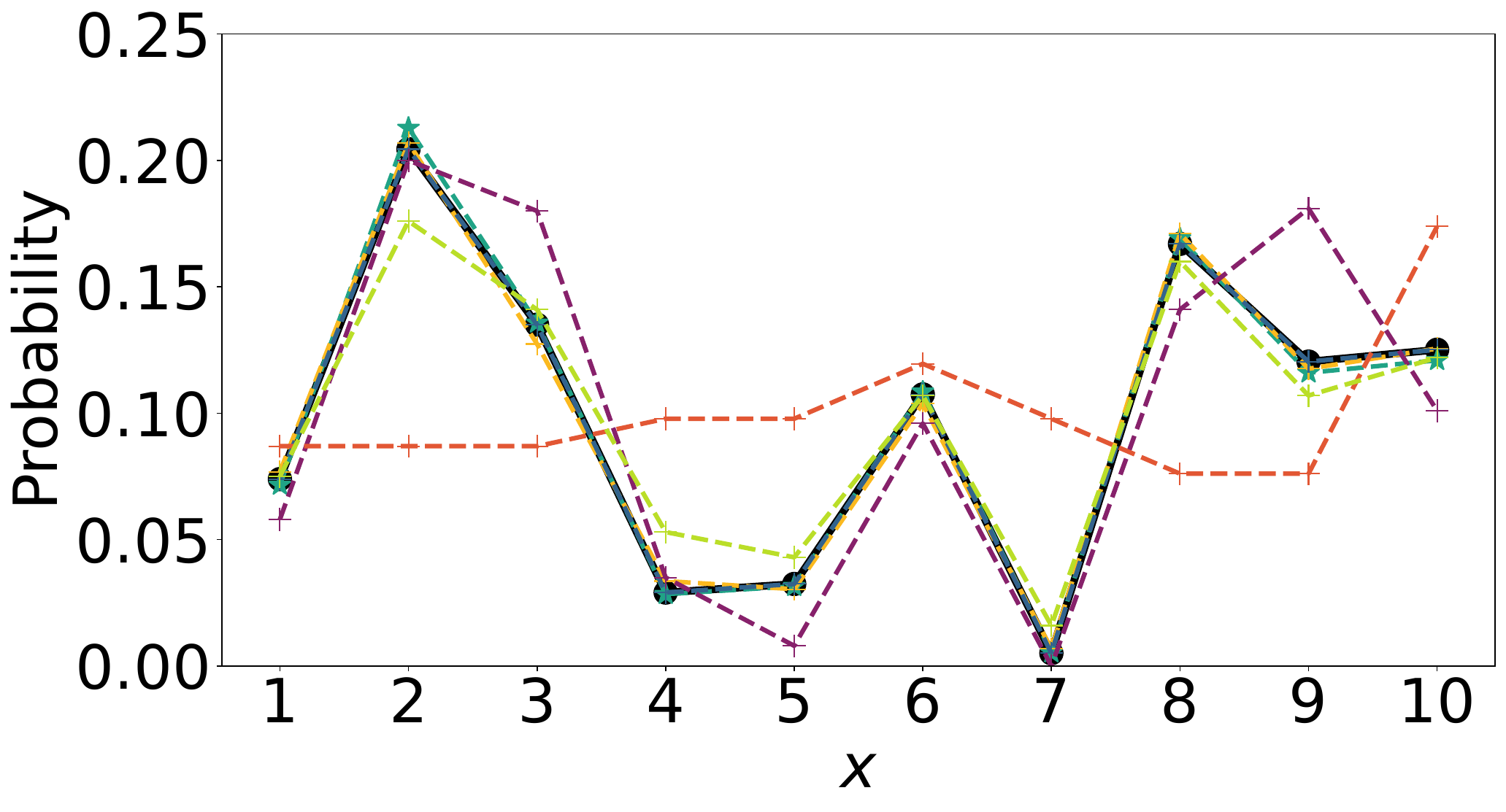}
        \caption{1D PMF}
        \label{fig:onedim_prbs}
    \end{subfigure}\\[1ex]
    \begin{subfigure}{0.45\linewidth}
        \centering
        \includegraphics[width=\linewidth]{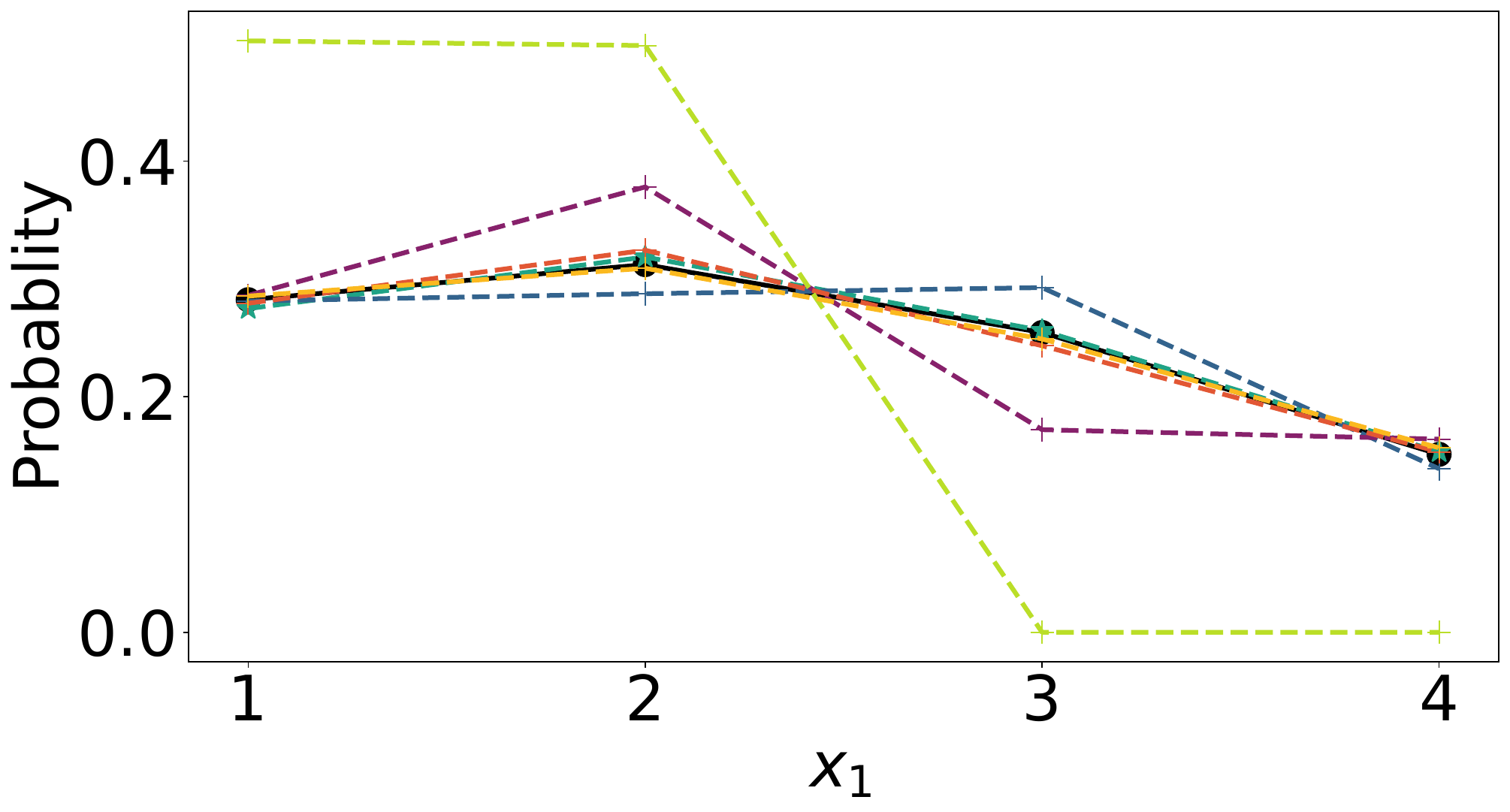}
        \caption{2D $x_1$-marginal PMF}
        \label{fig:twodim_prbs1}
    \end{subfigure}%
    \begin{subfigure}{0.45\linewidth}
        \centering
        \includegraphics[width=\linewidth]{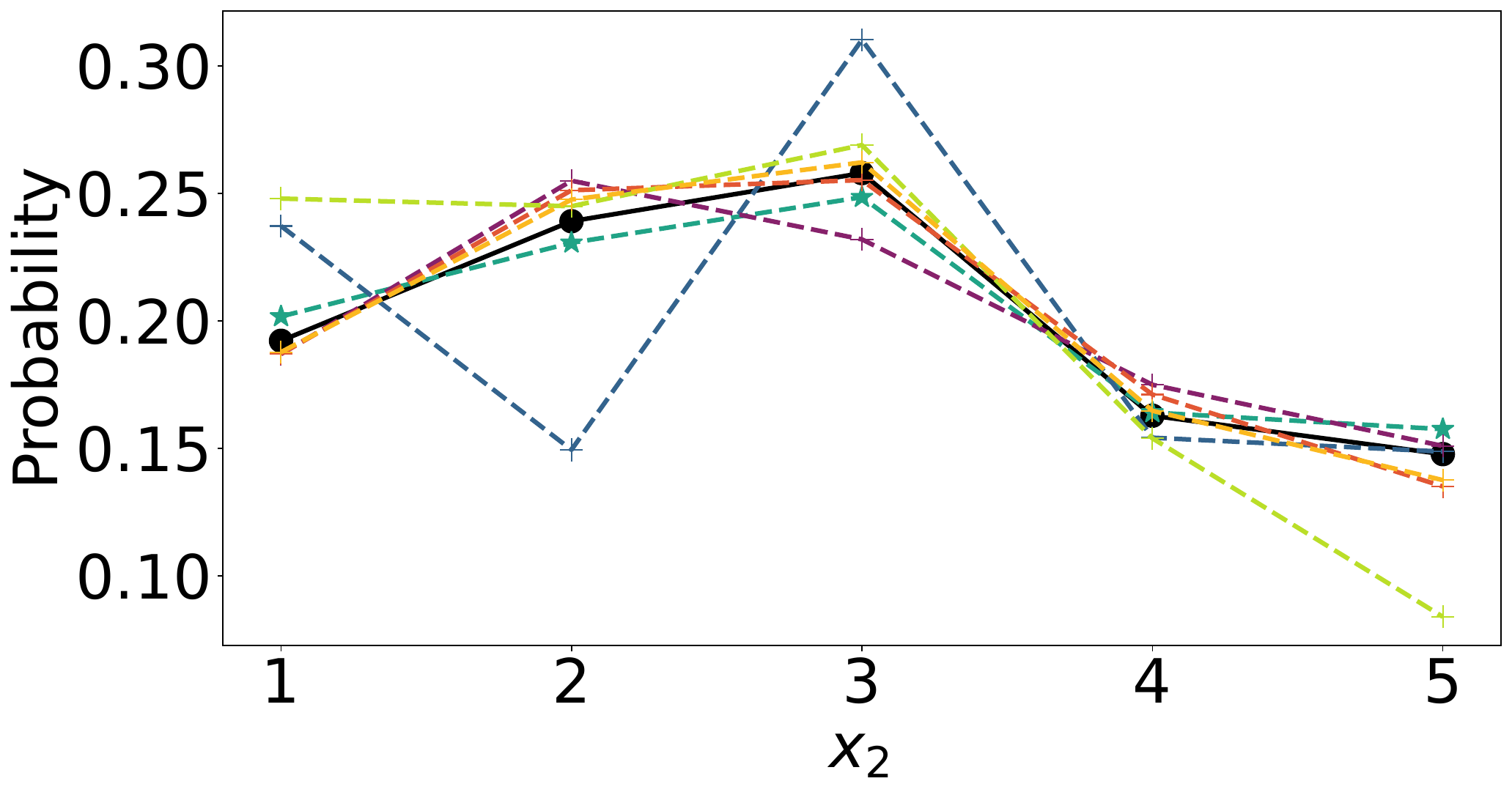}
        \caption{2D $x_2$-marginal PMF}
        \label{fig:twodim_prbs2}
    \end{subfigure}\\[1ex]
    \begin{subfigure}{0.45\linewidth}
        \centering
        \includegraphics[width=\linewidth]{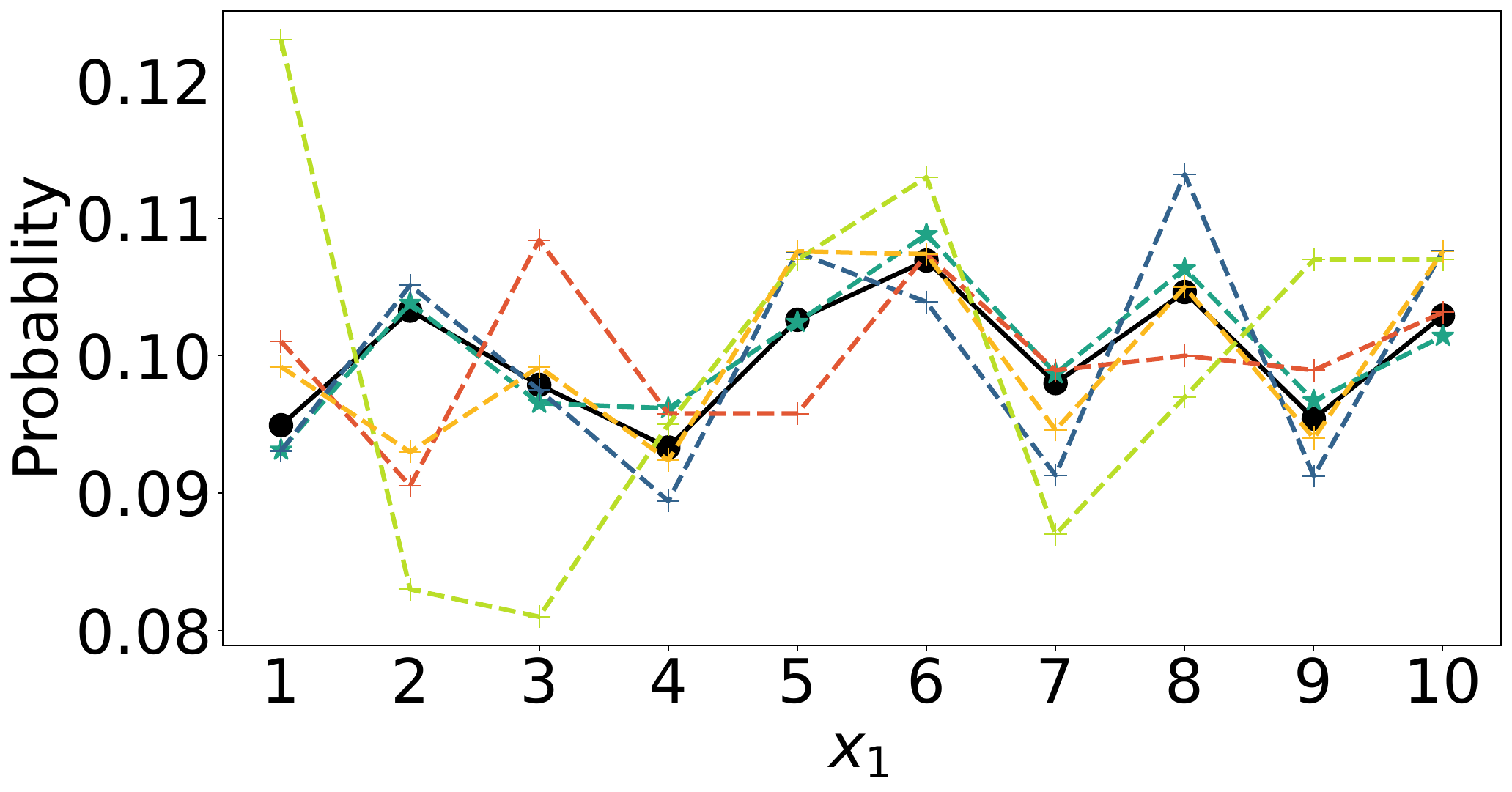}
        \caption{3D $x_1$-marginal PMF}
        \label{fig:threedim_prbs1}
    \end{subfigure}%
    \begin{subfigure}{0.45\linewidth}
        \centering
        \includegraphics[width=\linewidth]{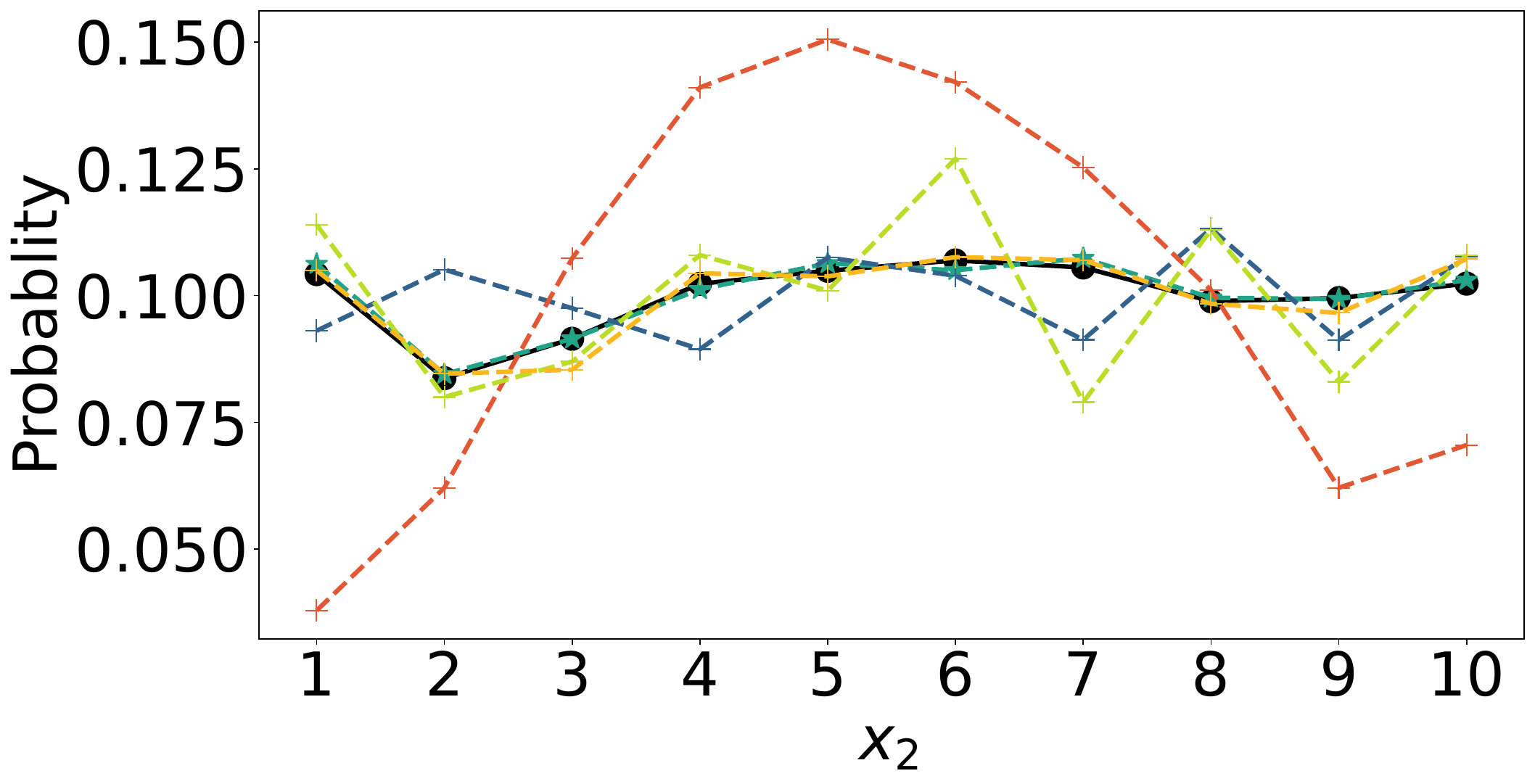}
        \caption{3D $x_2$-marginal PMF}
        \label{fig:threedim_prbs2}
    \end{subfigure}\\[1ex]
    \begin{subfigure}{0.45\linewidth}
        \centering
        \includegraphics[width=\linewidth]{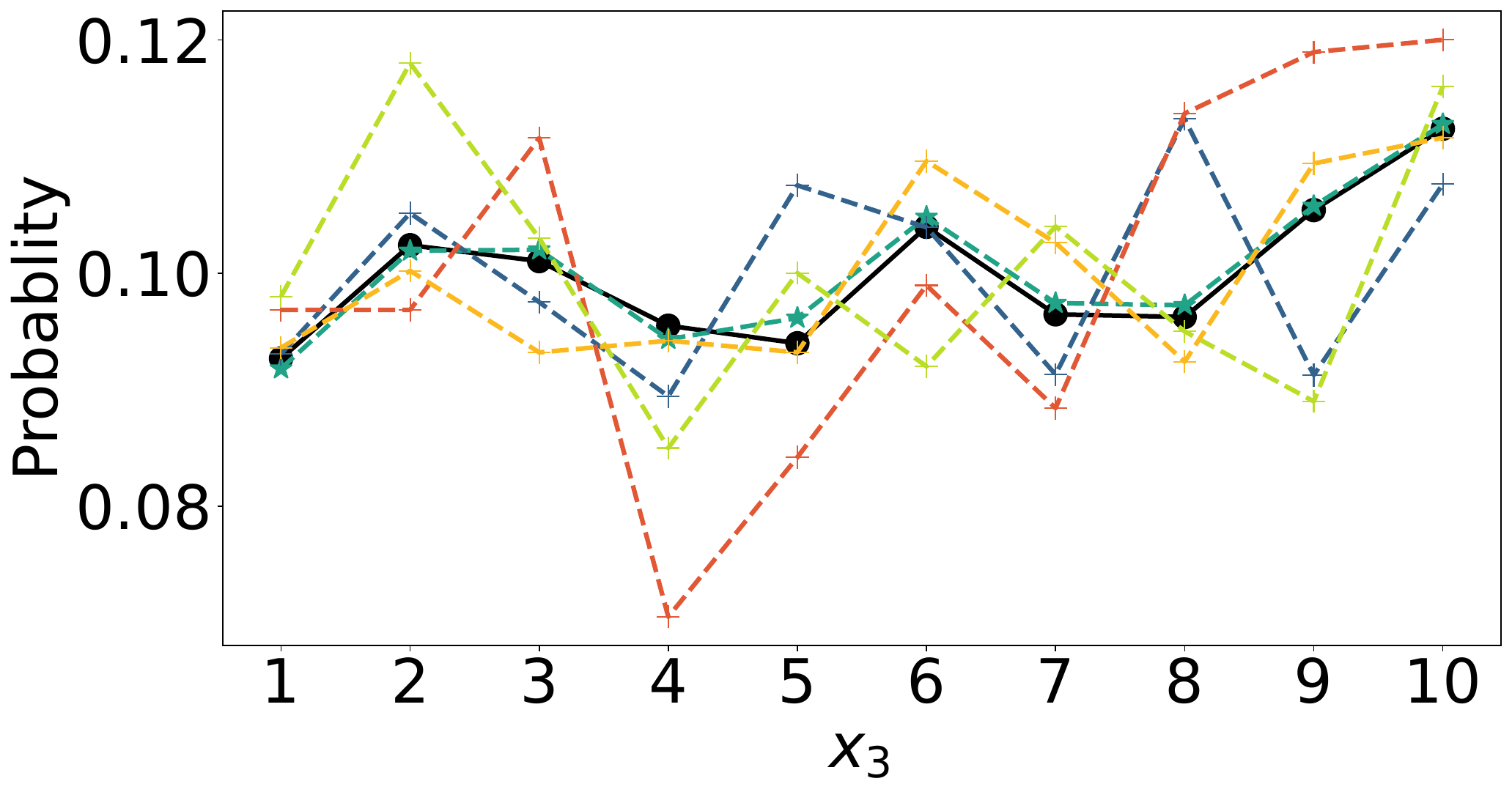}
        \caption{3D $x_3$-marginal PMF}
        \label{fig:threedim_prbs3}
    \end{subfigure}%
    \begin{subfigure}{0.45\linewidth}
        \centering
        \includegraphics[width=\linewidth]{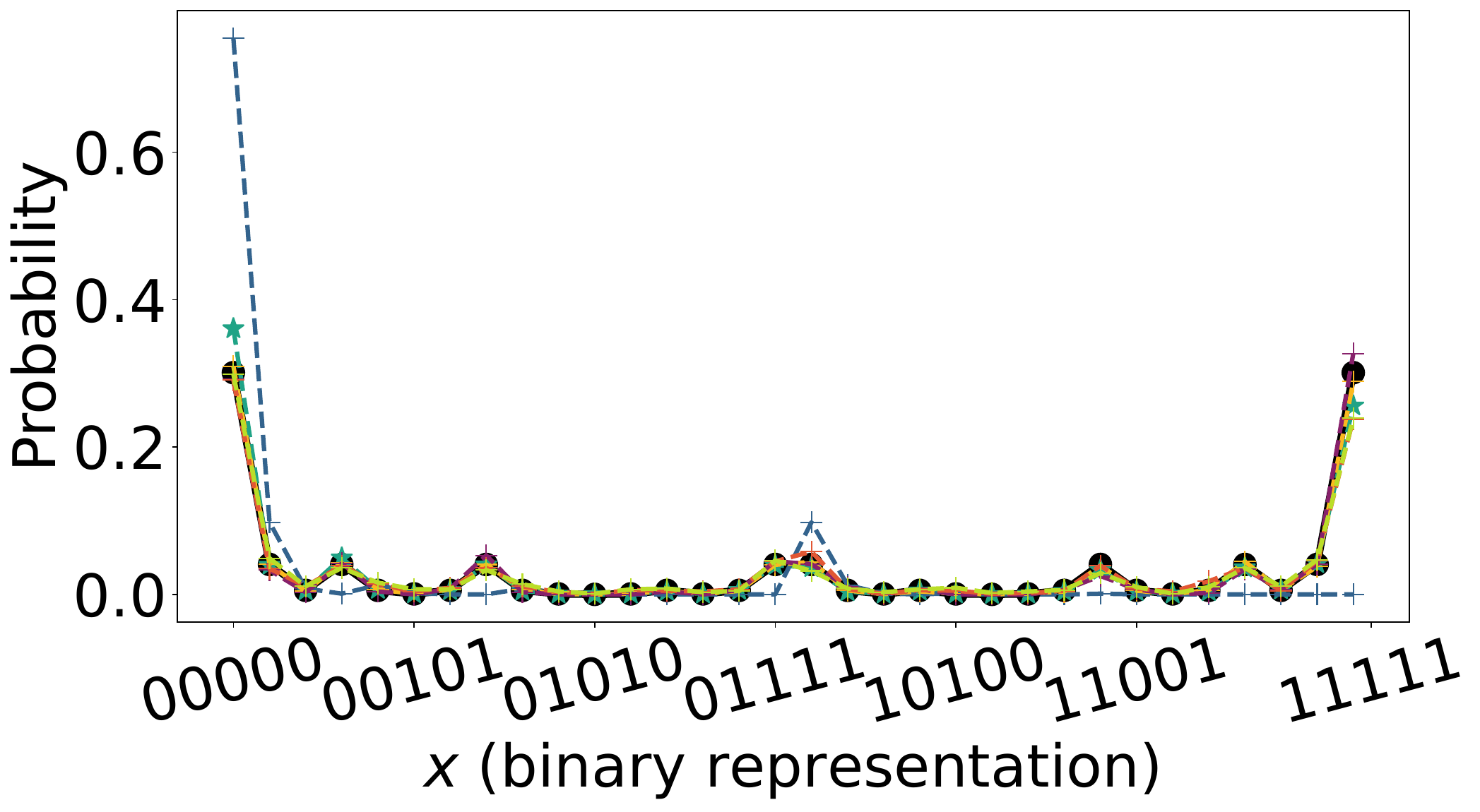}
        \caption{Ising $M=5$ flattened PMF}
        \label{fig:ising_prbs}
    \end{subfigure}
    \caption{True and approximated PMFs of the
    examples with tractable normalizing constant.
    Concrete-relaxed flows are not shown in the 3D example
    since even the optimal approximation---across the architecture search---is
    very poor.
    For the Ising model,
    we treat each $x\in\mcX$ as an element of $\{0,1\}^M$
    (i.e., a binary representation)
    and show the flattened PMF in ascending order of binary representation.
    The legend is shared across figures.}
    \label{fig:discrete_toy_examples}
\end{figure}
\section{MAD MIX KL-OPTIMAL WEIGHTING} \label{sec:kl_weighting}

One of the main benefits of MAD Mix over Gibbs sampling
and other stochastic MCMC samplers is having access to density estimates
of the approximation.
Combined with the availability of i.i.d. samples,
this allows us to estimate the ELBO and therefore 
to optimally weight two different MAD Mix flows, 
each initialized in a different reference distribution.
This is useful, for example,
when we know where certain high posterior mass regions are
but a usual MCMC sampler would struggle to mix between them.
Here we consider the case where there are two such regions,
but this can be extended to an arbitrary number.
Formally, we consider a variational proposal of the form
\[
    q_N=wq_{N,0}+(1-w)q_{N,1},
\]
where $w\in(0,1)$, $q_{N,i}=\frac{1}{N}\sum_{n=0}^{N-1} T_{\mathrm{MAD}}^n q_i$,
and $q_i$ is a reference distribution that covers the region $i$, $i=0,1$.
We then select
\[
    w=\arg\min_{\alpha\in(0,1)}
    \mathrm{D}_{\mathrm{KL}}
    (\alpha q_{N,0}+(1-\alpha)q_{N,1}\,||\,\pi),
\]
which can be estimated via gradient descent since
\[
    \frac{\mathrm{d}}{\mathrm{d}\alpha}
    \mathrm{D}_{\mathrm{KL}}
    (\alpha q_{N,0}+(1-\alpha)q_{N,1}\,||\,\pi)
    =\mathbb{E}_{q_{N,0}}
    \left[\log\frac{\alpha q_{N,0}+(1-\alpha)q_{N,1}}{\pi}\right]
    -\mathbb{E}_{q_{N,1}}
    \left[\log\frac{\alpha q_{N,0}+(1-\alpha)q_{N,1}}{\pi}\right].
\]
In practice,
one can generate samples from $q_{N,0}$ and $q_{N,1}$
and use them to estimate the expectations.

\section{ISING MODEL DETAILS} \label{sec:ising_details}

Recall that the Ising model has target density
\[
    \pi(x)\propto\exp\left\{\beta\sum_{m=1}^{M-1} x_m x_{m+1}\right\}
\]
with $\beta>0$ the inverse temperature
and $x\in\{-1,+1\}^M=\mcX$.
The full conditionals can be found in closed form
by analyzing only the terms affecting each particle's immediate neighbors.
For the particles $x_1$ and $x_M$ the full conditional
only depends on their single neighbor:
\[
    \pi_1(x)=\frac{\exp(\beta x x_2)}{2\cosh(\beta)},\qquad
    \pi_M(x)=\frac{\exp(\beta x x_{M-1})}{2\cosh(\beta)}.
\]
The normalizing constant is tractable since it involves adding two terms
and can be simplified since $\cosh(\beta)=\cosh(-\beta)$.
The probability for particles with two neighbors 
$x_m,  1<m<M$, is likewise given by
\[
    \pi_m(x)=\frac{\exp(\beta x (x_{m-1} + x_{m+1}))}{2\cosh(\beta (x_{m-1}+x_{m+1}))}.
\]

\section{GAUSSIAN MIXTURE MODEL EXPERIMENT DETAILS} \label{sec:gmm_fc}

We use the labels $x_{1:N}$ to rewrite the likelihood as
a product over the sample and label indices:
\[
    \ell(y_{1:N};x_{1:N},w_{1:K},\mu_{1:K},\Sigma_{1:K}) 
    = \prod_{n=1}^N \prod_{k=1}^K 
    (w_k \phi_{\mu_k,\Sigma_k}(y_n))^{\mathbb{I}(x_n=k)},
\]
where $\phi_{\mu,\Sigma}(y)$ is the density of a
$\mathcal{N}(\mu,\Sigma)$ distribution evaluated at $y\in\reals^D$.

We considered uninformative and independent prior distributions 
for each of the $NK+3K$ parameters:
\[
    X_n \given w &\sim \mathrm{Categorical}(w_1,\dots,w_K),\quad n\in[N],\\
    w&\sim\distDir(\alpha_1,\dots,\alpha_K),\\
    \mu_k \given \Sigma_k & \sim \mathcal{N}(m_{0,k},\Sigma_k),\quad k\in[K],\\
    \Sigma_k & \sim \distInvWish(S_{0,k},\nu_{0,k}), \quad k\in[K].
\]
We set $\alpha_k=1$ for all $k$,
reflecting little prior knowledge of the weights.
We also set $\nu_{0,k}=1$
and chose $m_{0,k}$ and $S_{0,k}$ 
by visually inspecting the data for all $k$.
We initialized MAD Mix, the Gibbs sampler,
and the mean-field algorithm at these values for fairness.

These prior distribution are conjugate and
the full conditionals can be found in closed form:
\[
    X_n \given w,\mu_{1:K},\Sigma_{1:K}&\sim\distCat(
        w_1\phi_{\mu_1,\Sigma_1}(y_n),\dots,w_K\phi_{\mu_K,\Sigma_K}(y_n)
        ),\quad n\in[N],\\
    w \given X_{1:N} &\sim 
    \distDir(\alpha_1+N_1,\dots,\alpha_K+N_K),\quad k\in[K],\\
    \mu_k \given \Sigma_k, X_{1:N} &\sim
    \distNorm(\bary_k,\Sigma_k/N_k),\quad k\in[K],\\
    \Sigma_k \given X_{1:N} &\sim
    \distInvWish(S_k,N_k-D-1),\quad k\in[K].
\]
Above, $D$ is the dimension of the data,
$N_k=\sum_n \ind(x_n=k)$ is the number of elements in cluster $k$,
$\bar{y}_k=N_k^{-1}\sum_n y_n\ind(x_n=k)$ is the mean of elements in cluster $k$,
and $S_k=\sum_n (y_n-\bar{y}_k)(y_n-\bar{y}_k)^\top \ind(x_n=k)$ the 
corresponding scaled covariance.
The mean-field algorithm also has closed-form updates;
we followed \citet[][Sec.~10.2]{bishop2006}.

\paragraph{MAD Mix implementation}

For the deterministic Hamiltonian move,
we need the score function of the parameters $(w,\mu_{1:K},\Sigma_{1:K})$.
Note that the score w.r.t. the weights will only depend on $p(w)$
and the score w.r.t. the $k$th mean will only depend on $p(\mu_k\,|\,\Sigma_k)$.

The score w.r.t. the weights is then
\[
    \nabla_w \log p(w,\mu,\Sigma)
    =\nabla_w \log p(w)
    =\nabla_w \sum_k N_k \log w_k
    =\left(\frac{N_1}{w_1},\dots,\frac{N_K}{w_K}\right)^\top.
\]
The score w.r.t. the $k$th mean is 
\[
    \nabla_{\mu_k} \log p(w,\mu,\Sigma)
    =\nabla_{\mu_k} \log p(\mu_k\,|\,\Sigma)
    =-N_k\Sigma_k^{-1}(\mu_k-\bar{y}_k).
\]
Finally, the score w.r.t. the $k$th covariance depends on both 
the mean PDF and the covariance PDF:
\[
    \nabla_{\Sigma_k} \log p(w,\mu,\Sigma)
    =\nabla_{\Sigma_k}\log p(\mu_k\,|\,\Sigma_k) + \log p(\Sigma_k).
\]
The first term is
\[
    \nabla_{\Sigma_k}\log p(\mu_k\,|\,\Sigma_k)
    =-\frac{1}{2}\Sigma_k^{-\top}-\frac{N_k}{2}(\mu_k-\bar{y}_k)(\mu_k-\bar{y}_k)^\top,
\]
where we used the identities $\partial \log|X|=X^{-\top}$
and $\partial a^\top X b=ab^\top$ \citep{petersen2008matrixcookbook}.
The second term is
\[
    \nabla_{\Sigma_k}\log p(\Sigma_k)
    =-\frac{N_k}{2}\Sigma_k^{-\top}+\frac{1}{2}\left(\Sigma_k^{-1}S_k\Sigma_k^{-1}\right)^\top,
\]
where we used the identity 
$\partial\mathrm{tr}(AX^{-1}B)=-X^{-\top}A^\top B^\top X^{-\top}$
\citep{petersen2008matrixcookbook}.
Adding together these two expressions yields the score w.r.t. the $k$th covariance:
\[ \label{eq:SigmaScore}
    \nabla_{\Sigma_k} \log p(w,\mu,\Sigma)
    =-\frac{1}{2}(1+N_k)\Sigma_k^{-\top}
    -\frac{N_k}{2}(\mu_k-\bar{y}_k)(\mu_k-\bar{y}_k)^\top
    +\frac{1}{2}\left(\Sigma_k^{-1}S_k\Sigma_k^{-1}\right)^\top.
\]
In practice, we avoid working with the covariance matrix directly 
since it has to be symmetric and positive definite,
and taking Hamiltonian steps can make the resulting matrix inadmissible.
Instead, given a covariance matrix $\Sigma$,
let $\Sigma=LL^\top$ be its (unique) Cholesky decomposition.
The matrix $L$ is lower triangular and has positive diagonal elements,
and hence we only need to store the $D+\binom{D}{2}$ non-zero elements.
To further remove the positiveness condition,
define $H$ as a copy of $L$ but with the diagonal log-transformed:
\[
    H_{ij}=\begin{cases}
        0,&i<j,\\
        \log L_{ij},&i=j,\\
        L_{ij},&i>j.
    \end{cases}
\]
By taking steps in $H$-space, 
we are guaranteed to get back a valid covariance matrix.
To map from $H$ to $\Sigma$,
we exp-transform the diagonal to get $L$ and then set $\Sigma=LL^\top$.
Let $f$ be the function that maps $\Sigma$ to $H$, so that $f(\Sigma)=H$.
The Jacobian of the transformation can be found in closed-form and
the resulting log density is
\[ \label{eq:Hlpdf}
    \log p(H)=\log p(\Sigma)+\sum_{d=1}^D (D-d+2)H_{dd} +D\log2,
\]
where $D$ is the dimension of observations 
\citep[][Theorem~4.2]{njoroge1988matrixjacobians}.

To take a Hamiltonian step in $H$ space, however,
we need $\nabla_H \log p(H)$.
We take the gradient w.r.t. $H$ in \cref{eq:Hlpdf}.
The third term does not depend on $H$.
The gradient of the second term is
a diagonal matrix with the $d$th diagonal entry equal to $D-d+2$.
Using the chain rule, the gradient of the first term is
\[
    \nabla_H \log p_\Sigma(f^{-1}(H))
    = \nabla_H f^{-1}(H)\nabla_\Sigma \log p_\Sigma(f^{-1}(H)).
\]
The second factor is equal to $\nabla_\Sigma \log p_\Sigma(\Sigma)$,
which we derived in \cref{eq:SigmaScore}, since $\Sigma=f(H)$.
The first factor is the Jacobian of the inverse transform $f$.
Note that this is a $D^2\times D^2$ matrix,
so the second factor should be understood as a $D^2$ vector (a flattened matrix).
We find this matrix by computing the Jacobian matrix of the transform
from $H$ to $L$, say $J_1$,
and multiplying it with the corresponding Jacobian
of the map from $L$ to $LL^\top=\Sigma$, $J_2$.
The first Jacobian matrix $J_1$ is diagonal with entries
either 1 or $\exp(H_{dd})$. Specifically,
\[
    J_1=\mathrm{Diag}
    (\exp(H_{11}),1,\dots,1,\exp(H_{22}),1,\dots,1,\exp(H_{DD}),1,\dots,1).
\]
For $J_2$, we introduce the $D^2\times D^2$ commutation matrix $K_D$.
If $A$ is a $D\times D$ matrix denote by 
$\mathrm{vec}(A)$ the vectorized or flattened version of $A$, i.e., 
the $\reals^{D^2}$ vector obtained by stacking the columns of $A$
one after the other.
Then $K_D$ satisfies that $K_D\mathrm{vec}(A)=\mathrm{vec}(A^\top)$ and we have
\[
    J_2=\frac{\partial\mathrm{vec}(LL^\top)}{\partial\mathrm{vec}(L)}
    =(L\otimes I_D)+(L\otimes I_D)K_D,
\]
where $\otimes$ is the Kroenecker product
and $I_D$ is the $D\times D$ identity matrix.
Finally, $\nabla_H f^{-1}(H)=J_2 J_1$.
We also use this decomposition for all the continuous-embedding flows.
\section{SPIKE-AND-SLAB MODEL DETAILS} \label{sec:sas_details}

We consider $N$ real-valued observations $y_{1:N}$ with accompanying
$P$ covariates $x_n\in\reals^P$, 
which we stack horizontally in a design matrix $X\in\reals^{N\times P}$.
We assume a linear regression setting where
\[
    y_n\sim\distNorm(\beta^\top x_n, \sigma^2),
\]
with regression coefficients $\beta\in\reals^P$ and unknown variance $\sigma^2$.
To induce sparseness in $\beta$,
we introduce additional latent variables $\theta\in(0,1)$,
$\gamma_{1:P}\in\{0,1\}^P$, and $\tau^2>0$, and
consider the following hierarchical model:
\[
    \tau^2&\sim\distInvGam(\alpha_1,\alpha_2),\\
    \sigma^2&\sim\distGam(\frac{1}{2},\frac{s^2}{2}),\\
    \theta&\sim\distBeta(a,b),\\
    \gamma_p&\sim\distCat(1-\theta, \theta),\qquad p=1,\dots,P,\\
    \beta_p&\sim (1-\gamma_p)\delta_0 + \gamma_p\distNorm(0,\sigma^2\tau^2),
    \qquad p=1,\dots,P,
\]
where $\delta_0$ is a Dirac delta at 0.
The variance of the regression coefficients $\nu^2=\sigma^2\tau^2$
depends on both the variance of the observations $\sigma^2$ and on $\tau^2$
to allow the prior to scale with the scale of the observations.
We set $\alpha_1=\alpha_2=0.1$, $s^2=0.5$, and $a=b=1$
to reflect little prior knowledge about the latent variables.
We initialized the regression coefficients in 
MAD Mix, Gibbs sampling, and mean-field VI
at the least-squares estimator.

These prior distributions are conjugate and
the full conditionals can be found in closed form:
\[
    \tau^2\given\sigma^2,\beta,\gamma_{1:P},
    &\sim\distInvGam\left(\frac{1}{2}+\frac{1}{2}\sum_{p=1}^P\gamma_p,
    \frac{s^2}{2}+\frac{\beta^\top\beta}{2\sigma^2}\right),\\
    \sigma^2\given y,\beta
    &\sim\distGam\left(\alpha_1 + \frac{N}{2},
    \alpha_2+\frac{1}{2}(y-X\beta)^\top(y-X\beta)\right),\\
    \theta\given\gamma_{1:P} 
    &\sim\distBeta\left(a+\sum_{p=1}^P\gamma_p, b+P-\sum_{p=1}^P\gamma_p \right),\\
    \beta\given y,\gamma_{1:P},\tau^2,\sigma^2
    &\sim\distNorm\left(\frac{1}{\sigma^2}HX^\top y, H\right),\\
    \gamma_p \given \beta,\theta,\tau^2,y
    &\sim\distCat(1-\xi_p,\xi_p),
\]
where $H=\sigma^2(X^\top X + \frac{1}{\tau^2}I_P)^{-1}$
is the projection matrix in ridge regression
(sans an additional $X^\top$ term). 
The Categorical probabilities for the full conditional of $\gamma_p$ are given by
\[
    1-\xi_p=
    \frac{1-\theta}
    {\tau^{-1}
    \exp\left(\frac{\left(\sum_{n=1}^N x_n^\top z_p \right)^2}
    {2\sigma^2\left(\sum_{n=1}^N x_n^\top x_n + \frac{1}{\tau^2}\right)}\right)
    \left(\sum_{n=1}^N x_n^\top x_n + \frac{1}{\tau^2}\right)^{-1/2}
    \theta + (1-\theta)},
\]
where $z_p = y-X\beta_{-p}$ are the residuals from the model with
parameters $\beta_{-p}$ in which we set $\beta_p=0$.
The derivations can be found in \citet{dablanderSAS}.
For mean-field VI, we followed \citet{ray2022sasmf}
and used their software \citep{sparsevb}.

\paragraph{MAD Mix implementation}
In practice, we transform the restricted variables $\theta\in(0,1),\sigma^2,\tau^2>0$
into real-valued parameters to prevent the Hamiltonian dynamics from 
resulting in unfeasible values. Specifically, we reparametrize
\[
    \theta_u = \log \frac{\theta}{1-\theta},\quad
    \tau_u^2 = \log \tau^2,\quad
    \sigma_u^2 = \log \sigma^2.
\]
These transformations are bijective and have Jacobians
\[
    J_{\theta_u} = \frac{1}{\theta(1-\theta)},\quad
    J_{\tau_u^2} = \frac{1}{\tau^2},\quad
    J_{\sigma_u^2} = \frac{1}{\sigma^2}.
\]
We do inference over the unrestricted parameters
and also use them for all continuous-embedding flows.

As with the GMM experiment,
we need the score functions of the continuous variables for
the Hamiltonian step in MAD Mix.
Let $\pi$ denote the spike-and-slab posterior distribution
as a function of the continuous variables only
(see \cref{subsec:mad_extension} for more details).
Then the score functions are given by
\[
    \grad_{\theta_u} \log \pi(\theta_u,\sigma_u^2,\tau_u^2,\beta; \gamma_{1:P})
    &= \frac{a-1}{\theta}-\frac{b-1}{\theta}-\frac{1}{\theta(1-\theta)},\\
    \grad_{\tau_u^2} \log \pi(\theta_u,\sigma_u^2,\tau_u^2,\beta; \gamma_{1:P})
    &= \frac{s^2}{2(\tau^2)^2} - \frac{1+\sum_{p=1}^P \gamma_p}{2\tau^2}
    +\frac{1}{2\sigma^2\tau^2}\sum_{p=1}^P \gamma_p \beta_p^2,\\
    \grad_{\sigma_u^2} \log \pi(\theta_u,\sigma_u^2,\tau_u^2,\beta; \gamma_{1:P})
    &= \frac{2\alpha_2 + (y-X\beta)^\top(y-X\beta)}{2(\sigma^2)^2}
    -\frac{2\alpha_1+N+\sum_{p=1}^P \gamma_p}{2\sigma^2}
    +\frac{1}{2\sigma^2\tau^2}\sum_{p=1}^P \gamma_p \beta_p^2,\\
    \grad_{\beta} \log \pi(\theta_u,\sigma_u^2,\tau_u^2,\beta; \gamma_{1:P})
    &= \frac{1}{\sigma^2}\left(X^\top y - X^\top X\beta - \frac{1}{\tau^2}\beta\right).
\]
\clearpage
\section{PROOFS} \label{sec:proofs}

First we state and prove a series of results
that will be helpful in the proof of \cref{prop:fc_mp}.

\subsection{Derivative under modulo operation}
To calculate the density of the pushfoward under $T_{\mathrm{MAD}}$,
we need to calculate the derivative of the modulo operation.
\cref{lem:lcg_der} below shows that the update $\rho\mapsto\rho'$
in Step (2) of the MAD map has unit derivative.
\blem\label{lem:lcg_der}
  Let $\rho\in(0,1)$ and define $\rho'=\rho+\xi \mod 1$, where $\xi\in\reals$.
  Then the transformation $\rho\mapsto\rho'$ has unit derivative:
  \[
    \der{\rho'}{\rho}=1,\qquad\text{for }\rho\text{ s.t. }\rho+\xi\mod1\neq0.
  \]
  If $\rho+\xi\mod1=0$ then the derivative is not defined.
\elem

\bprfof{\cref{lem:lcg_der}}
  We can rewrite the modulo function as the original value
  minus the integer part of the remainder:
  \[
    \rho+\xi\mod 1=\rho+\xi-\lfloor \rho+\xi-1\rfloor.
  \]
  When taking derivative w.r.t. $\rho$,
  the term $\rho+\xi$ has unit derivative
  and the floor term has derivative zero since 
  the floor function is piecewise constant in $(0,1)$.
\eprfof

\subsection{Change of variables for joint discrete and continous transformations}
The usual change of variables formula is valid when
all the variables are continuous (e.g., \cref{eq:nfs_density})
or when all the variables are discrete (in which case there is no Jacobian term).
In this section, we develop an analogue for the case where some variables are discrete
and some are continuous.
We consider the setting where a base distribution $\pi_0$
has exactly one discrete and one continuous component,
but the result generalizes.
Formally, let $\pi_0$ be a density on $\nats\times\reals$,
$T:\nats\times\reals\to\nats\times\reals$ be an invertible transformation
where $T(n,x)=(T_\mathrm{d}(n,x),T_\mathrm{c}(n,x))$,
and $\pi_1=T\pi_0$ be the pushforward of $\pi_0$ under $T$.
Throughout, we consider two assumptions:
\bitems
  \item[(A1)]\label{a1} The map $T$ is invertible.
  \item[(A1)]\label{a2} For all $n\in\nats$,
  the map $T_\mathrm{c}(n,\cdot)$ is invertible.
\eitems

First we introduce some notation that will be useful to prove the main result.
\blem\label{lem:AnmBnm}
Define 
\[
  A_{nm} = \left\{x\in\reals\given T_\mathrm{d}(n,x)=m\right\},\qquad
  B_{nm}=\{y\in\reals\given y=T_\mathrm{c}(n,x),\,x\in A_{nm}\},\qquad
  n,m\in\nats.
\]
Then, under Assumptions (A1) and (A2),
for any fixed $n\in\nats$ the $(A_{nm})_{m=1}^\infty$ are a partition of $\reals$
and, for any fixed $m\in\nats$, the $(B_{nm})_{n=1}^\infty$ are a partition of $\reals$.
\elem

\bprfof{\cref{lem:AnmBnm}}
For all $n\in\nats$, each $x\in\reals$ is in one of the sets $(A_{nm})_{m\in\nats}$:
the one for which $T_\mathrm{d}(n,x)=m$.
Therefore for any fixed $n$ the $(A_{nm})_m$ are disjoint and cover all of $\reals$.

To show disjointness of the $(B_{nm})_n$,
suppose that $y\in B_{nm}\cap B_{n'm}$ for $n\neq n'$.
Then by definition
there would exist $x\in A_{nm},x'\in A_{n'm}$ such that
$T_\mathrm{c}(n,x)=y=T_\mathrm{c}(n',x')$.
But since $x,x'$ belong to $A_{nm}$ and $A_{n'm}$ (respectively),
$T_\mathrm{d}(n,x)=m=T_\mathrm{d}(n',x')$.
Hence $T(n,x)=T(n',x')$ and thus $(n,x)=(n',x')$ because $T$ is invertible,
which is a contradiction since we assumed $n\neq n'$.
To show that the $(B_{nm})_n$ cover $\reals$ for a fixed $m$,
given $y\in\reals$ let $(n,x)=T^{-1}(m,y)$, i.e.,
$T_\mathrm{d}(n,x)=m$ and $T_\mathrm{c}(n,x)=y$.
The first equality implies $x\in A_{nm}$, 
which together with the second equality shows $y\in B_{nm}\subseteq\cup_n B_{nm}$.
\eprfof

Now we state the main result.
\bprop\label{prop:cov}
  Let $J_\mathrm{c}$ be the conditional Jacobian w.r.t. the continous variable $x$
  of the map $T_\mathrm{c}(n,\cdot)$,
  and let $J_\mathrm{c}^{-1}$ be the Jacobian of the inverse map:
  \[
    J_\mathrm{c}(n,x)
    :=\der{T_\mathrm{c}(n,\cdot)}{x}\Bigg|_{x},\qquad
    J_\mathrm{c}^{-1}(n,x)
    :=\der{T_\mathrm{c}^{-1}(n,\cdot)}{y}\Bigg|_{y=T_\mathrm{c}(n,x).}
  \]
  Then under Assumptions (A1) and (A2),
  the density of the pushforward $\pi_1$
  is given by the following change of variables formula:
  \[
    \pi_1(m,y)=\pi_0(T^{-1}(m,y))\left|J_\mathrm{c}^{-1}(T^{-1}(m,y))\right|,\qquad
    m\in\nats,y\in\reals.
  \]
\eprop

\bprfof{\cref{prop:cov}}
Consider an arbitrary measurable,
absolutely integrable function $f:\nats\times\reals\to\reals$.
By the change of variables theorem
\citep[][Theorem~5.2]{cinlar2011probability},
\[
  \EE_{\pi_1}[f(M,Y)]
  = \sum_{n=1}^{\infty}\int_{\reals}\pi_0(n,x)f(T(n,x))\,\dee x.
\]
By the Radon-Nikodym theorem
\citep[][Theorem~5.11]{cinlar2011probability},
the left-hand side can also be expressed as an integral
w.r.t. the pushforward $\pi_1$.
Our strategy will be to rewrite the right-hand side into
an integral containing the density in the statement of \cref{prop:cov}
and then use the almost-sure uniqueness of the Radon-Nikodym derivative.

Since the subsets $(A_{nm})_{m\in\nats}$
are a partition of $\reals$ for all $n\in\nats$
by \cref{lem:AnmBnm},
we can rewrite the inner integral by summing over the subsets
and using the fact that in $A_{nm}$ we have $T_\mathrm{d}(n,x)=m$:
\[
  \int_{\reals}\pi_0(n,x)f(T(n,x))
  =\sum_{m=1}^{\infty}\int_{A_{nm}}\pi_0(n,x)f(m,T_\mathrm{c}(n,x))\,\dee x,
  \quad n\in\nats.
\]
It is possible that some $A_{nm}=\emptyset$.
This is not a problem since the (Lebesgue) integral over an empty set is 0.
Next we do the following change of variables: $y=T_\mathrm{c}(n,x)$.
This is well-defined by Assumption (A2),
and specifically we have $x=T_\mathrm{c}^{-1}(n,y)$,
where we are inverting w.r.t. the continuous variable for fixed $n$.
Hence the Jacobian term is $\dee x=|J_{\mathrm{c}}(n,x)|\dee y$.
Furthermore, note that $(n,x)=T^{-1}(m,y)$ by Assumption (A1)
and that the integration domain is now $B_{nm}$.
Thus the integral over $A_{nm}$ is
\[
  \int_{A_{nm}}\pi_0(n,x)f(m,T_\mathrm{c}(n,x))\,\dee x
  =\int_{B_{nm}}\pi_0(T^{-1}(m,y))
  \left|J_{\mathrm{c}}(T^{-1}(m,y))\right|f(m,y)\,\dee y,
  \quad n\in\nats.
\]
We plug in this expression into the 
expectation w.r.t. $\pi_1$
and change the order of the sums by Fubini's theorem:
\[
  \EE_{\pi_1}[f(M,Y)]
  =\sum_{m=1}^{\infty}\sum_{n=1}^{\infty}\int_{B_{nm}}\pi_0(T^{-1}(m,y))
  \left|J_{\mathrm{c}}(T^{-1}(m,y))\right|f(m,y)\,\dee y.
\]
The $(B_{nm})_n$ are a partition for each $m$ by \cref{lem:AnmBnm} so
we simplify the integral on the right-hand side into an integral over $\reals$:
\[
  \EE_{\pi_1}[f(M,Y)]
  =\sum_{m=1}^{\infty}\int_{\reals}\pi_0(T^{-1}(m,y))
  \left|J_{\mathrm{c}}(T^{-1}(m,y))\right|f(m,y)\,\dee y.
\]
Then by the almost-sure uniqueness of the Radon-Nikodym derivative
\citep[][Theorem~5.11]{cinlar2011probability},
\[
  \pi_1(m,y)=\pi_0(T^{-1}(m,y))\left|J_{\mathrm{c}}(T^{-1}(m,y))\right|
\]
for all $m\in\nats$ and for Lebesgue-almost all $y\in\reals$.
\eprfof

\subsection{Proof of \cref{prop:fc_mp}}
Now we have the necessary results to prove \cref{prop:fc_mp}.

\bprfof{\cref{prop:fc_mp}}
As discussed in \cref{subsec:theory}, the map $T_m$ is invertible.
Furthermore, the continuous restriction $u_m\mapsto u_m'$ 
is also invertible (for fixed $x_m$)
since it is a combination of non-zero affine transformations.
We assume w.l.o.g. that $u_m\in[0,1]$,
which in practice will be the case by construction.
Then by \cref{prop:cov}
and the inverse function rule
the density of $(x_m',u_m')$ is
\[
  \mathrm{Pr}(x_m',u_m')
  =\pi(x_m,u_m)\left|\der{u_m}{u_m'}\right|
  =\pi(x_m,u_m)\left|\der{u_m'}{u_m}\right|^{-1}
  =\tdpi_m(x_m)\left|\der{u_m'}{u_m}\right|^{-1}.
\]
We obtain the Jacobian term by manipulating the expression for $u_m'$
and using \cref{lem:lcg_der} and the Jacobian 
of $\rho_m$ w.r.t. $u_m$ from Step (1) in \cref{sec:main_idea}:
\[
  \der{u_m'}{u_m}
  &=\der{}{u_m}\left(
    \frac{\tdrho_m
    -F_m(x_m'-1)}{\tdpi_m(x_m')}
    \right)\\
  &=\frac{1}{\tdpi_m(x_m')}\der{\tdrho_m}{u_m}\\
  &=\frac{1}{\tdpi_m(x_m')}\der{\tdrho_m}{\rho_m}\der{\rho_m}{u_m}\\
  &=\frac{\tdpi_m(x_m)}{\tdpi_m(x_m')}.
\]
Plugging back into our previous result and
reconstructing $\tdpi$ from $\tdpi_m$ since $u_m'\in[0,1]$ yields the result:
\[
  \mathrm{Pr}(x_m',u_m')
  =\tdpi_m(x_m)\frac{\tdpi_m(x_m')}{\tdpi_m(x_m)}
  =\tdpi_m(x_m')\ind_{[0,1]}(u_m')
  =\tdpi(x_m',u_m').
\]
\eprfof

\subsection{Proof of \cref{prop:slice_mp}}
\bprfof{\cref{prop:slice_mp}}
Let $f:\mcX_1\times\mcX_2\to\mcX_1\times\mcX_2$ be any measurable function.
We  show that the integral of $f$ w.r.t. $\pi$
and w.r.t. the pushforward $T\pi$ is the same
by using the disintegration of $\pi$ and appealing to
the measure-preserving property of $T_{x_1}$. Formally,
\[
  \int f(x_1,x_2) \pi(\dee x_1,\dee x_2)
  &=\int_{\mcX_1}\int_{\mcX_2}f(x_1,x_2)\pi_{2|1}(\dee x_2, x_1)\pi_1(\dee x_1)\\
  &=\int_{\mcX_1}\int_{\mcX_2}f(x_1,x_2)(T_{x_1}\pi_{2|1})(\dee x_2)\pi_1(\dee x_1)\\
  &= \int f(x_1,x_2)(\text{Id},T_{x_1})\pi(\dee x_1, \dee x_2).
\]
This shows that
\[
  \int f\,\dee\pi=\int f\,\dee T\pi
\]
for any measurable $f$, i.e., that $T$ is $\pi$-measure-preserving.
\eprfof

\end{document}